
%
\expandafter\ifx\csname phyzzx\endcsname\relax
 \message{It is better to use PHYZZX format than to
          \string\input\space PHYZZX}\else
 \wlog{PHYZZX macros are already loaded and are not
          \string\input\space again}%
 \endinput \fi
\catcode`\@=11 
\let\rel@x=\relax
\let\n@expand=\relax
\def\pr@tect{\let\n@expand=\noexpand}
\let\protect=\pr@tect
\let\gl@bal=\global
%
%
%
\newfam\cpfam
\newdimen\b@gheight             \b@gheight=12pt
\newcount\f@ntkey               \f@ntkey=0
\def\f@m{\afterassignment\samef@nt\f@ntkey=}
\def\samef@nt{\fam=\f@ntkey \the\textfont\f@ntkey\rel@x}
\def\setstr@t{\setbox\strutbox=\hbox{\vrule height 0.85\b@gheight
                                depth 0.35\b@gheight width\z@ }}
%
%
%
%
%

\font\fourteenrm=cmr10 scaled\magstep2
\font\twelverm=cmr10 scaled\magstep1
\font\tenrm=cmr10

\font\eightrm=cmr8
\font\sevenrm=cmr7
\font\sixrm=cmr6
\font\fiverm=cmr5
%
%

\font\fourteenbf=cmbx10 scaled\magstep2
\font\twelvebf=cmbx10 scaled\magstep1
\font\tenbf=cmbx10

\font\eightbf=cmbx8
\font\sevenbf=cmbx7
\font\sixbf=cmbx6
\font\fivebf=cmbx5
%
%
\font\seventeeni=cmmi10 scaled\magstep3     \skewchar\seventeeni='177
\font\fourteeni=cmmi10 scaled\magstep2      \skewchar\fourteeni='177
\font\twelvei=cmmi10 scaled\magstep1        \skewchar\twelvei='177
\font\teni=cmmi10 			    \skewchar\teni='177
\font\ninei=cmmi9			    \skewchar\ninei='177
\font\eighti=cmmi8                          \skewchar\eighti='177
\font\seveni=cmmi7                          \skewchar\seveni='177
\font\sixi=cmmi6                            \skewchar\sixi='177
\font\fivei=cmmi5			    \skewchar\fivei='177%
%
%
\font\seventeensy=cmsy10 scaled\magstep3     \skewchar\seventeensy='60
\font\fourteensy=cmsy10 scaled\magstep2      \skewchar\fourteensy='60
\font\twelvesy=cmsy10 scaled\magstep1        \skewchar\twelvesy='60
\font\tensy=cmsy10 			    \skewchar\tensy='60
\font\ninesy=cmsy9			    \skewchar\ninesy='60
\font\eightsy=cmsy8                          \skewchar\eightsy='60
\font\sevensy=cmsy7                          \skewchar\sevensy='60
\font\sixsy=cmsy6                            \skewchar\sixsy='60
\font\fivesy=cmsy5			    \skewchar\fivesy='60
%
%

\font\fourteenex=cmex10 scaled\magstep2
\font\twelveex=cmex10 scaled\magstep1
\font\tenex=cmex10
%
%

\font\fourteensl=cmsl10 scaled\magstep2
\font\twelvesl=cmsl10 scaled\magstep1
\font\tensl=cmsl10

\font\eightsl=cmsl8
%
%

\font\fourteenit=cmti10 scaled\magstep2
\font\twelveit=cmti10 scaled\magstep1
\font\tenit=cmti10

\font\eightit=cmti8

%
%
\font\fourteentt=cmtt10 scaled\magstep2
\font\twelvett=cmtt10 scaled\magstep1
\font\tentt=cmtt10

%
%
\font\fourteencp=cmcsc10 scaled\magstep2
\font\twelvecp=cmcsc10 scaled\magstep1
\font\tencp=cmcsc10
%
%

%
%

%
%
%

%
\def\fourteenf@nts{\relax
    \textfont0=\fourteenrm          \scriptfont0=\tenrm
      \scriptscriptfont0=\sevenrm
    \textfont1=\fourteeni           \scriptfont1=\teni
      \scriptscriptfont1=\seveni
    \textfont2=\fourteensy          \scriptfont2=\tensy
      \scriptscriptfont2=\sevensy
    \textfont3=\fourteenex          \scriptfont3=\twelveex
      \scriptscriptfont3=\tenex
    \textfont\itfam=\fourteenit     \scriptfont\itfam=\tenit
    \textfont\slfam=\fourteensl     \scriptfont\slfam=\tensl
    \textfont\bffam=\fourteenbf     \scriptfont\bffam=\tenbf
      \scriptscriptfont\bffam=\sevenbf
    \textfont\ttfam=\fourteentt
    \textfont\cpfam=\fourteencp }
\def\twelvef@nts{\relax
    \textfont0=\twelverm          \scriptfont0=\eightrm
      \scriptscriptfont0=\sixrm
    \textfont1=\twelvei           \scriptfont1=\eighti
      \scriptscriptfont1=\sixi
    \textfont2=\twelvesy           \scriptfont2=\eightsy
      \scriptscriptfont2=\sixsy
    \textfont3=\twelveex          \scriptfont3=\tenex
      \scriptscriptfont3=\tenex
    \textfont\itfam=\twelveit     \scriptfont\itfam=\eightit
    \textfont\slfam=\twelvesl     \scriptfont\slfam=\eightsl
    \textfont\bffam=\twelvebf     \scriptfont\bffam=\eightbf
      \scriptscriptfont\bffam=\sixbf
    \textfont\ttfam=\twelvett
    \textfont\cpfam=\twelvecp }
\def\tenf@nts{\relax
    \textfont0=\tenrm          \scriptfont0=\sevenrm
      \scriptscriptfont0=\fiverm
    \textfont1=\teni           \scriptfont1=\seveni
      \scriptscriptfont1=\fivei
    \textfont2=\tensy          \scriptfont2=\sevensy
      \scriptscriptfont2=\fivesy
    \textfont3=\tenex          \scriptfont3=\tenex
      \scriptscriptfont3=\tenex
    \textfont\itfam=\tenit     \scriptfont\itfam=\seveni  
    \textfont\slfam=\tensl     \scriptfont\slfam=\sevenrm 
    \textfont\bffam=\tenbf     \scriptfont\bffam=\sevenbf
      \scriptscriptfont\bffam=\fivebf
    \textfont\ttfam=\tentt
    \textfont\cpfam=\tencp }
%
%
\def\rm{\n@expand\f@m0 }
\def\mit{\n@expand\f@m1 }         
\def\cal{\n@expand\f@m2 }
\def\it{\n@expand\f@m\itfam}
\def\sl{\n@expand\f@m\slfam}
\def\bf{\n@expand\f@m\bffam}
\def\tt{\n@expand\f@m\ttfam}
\def\caps{\n@expand\f@m\cpfam}    
\def\em@{\rel@x\ifnum\f@ntkey=0 \it \else
        \ifnum\f@ntkey=\bffam \it \else \rm \fi \fi }
\def\em{\n@expand\em@}
\def\fourteenpoint{\fourteenf@nts \samef@nt \b@gheight=14pt \setstr@t }
\def\twelvepoint{\twelvef@nts \samef@nt \b@gheight=12pt \setstr@t }
\def\tenpoint{\tenf@nts \samef@nt \b@gheight=10pt \setstr@t }
\normalbaselineskip = 20pt plus 0.2pt minus 0.1pt
\normallineskip = 1.5pt plus 0.1pt minus 0.1pt
\normallineskiplimit = 1.5pt
\newskip\normaldisplayskip
\normaldisplayskip = 20pt plus 5pt minus 10pt
\newskip\normaldispshortskip
\normaldispshortskip = 6pt plus 5pt
\newskip\normalparskip
\normalparskip = 6pt plus 2pt minus 1pt
\newskip\skipregister
\skipregister = 5pt plus 2pt minus 1.5pt
\newif\ifsingl@
\newif\ifdoubl@
\newif\iftwelv@  \twelv@true
\def\singlespace{\singl@true\doubl@false\spaces@t}
\def\doublespace{\singl@false\doubl@true\spaces@t}
\def\normalspace{\singl@false\doubl@false\spaces@t}
\def\Tenpoint{\tenpoint\twelv@false\spaces@t}
\def\Twelvepoint{\twelvepoint\twelv@true\spaces@t}
\def\spaces@t{\rel@x
      \iftwelv@ \ifsingl@\subspaces@t3:4;\else\subspaces@t1:1;\fi
       \else \ifsingl@\subspaces@t3:5;\else\subspaces@t4:5;\fi \fi
      \ifdoubl@ \multiply\baselineskip by 5
         \divide\baselineskip by 4 \fi }
\def\subspaces@t#1:#2;{
      \baselineskip = \normalbaselineskip
      \multiply\baselineskip by #1 \divide\baselineskip by #2
      \lineskip = \normallineskip
      \multiply\lineskip by #1 \divide\lineskip by #2
      \lineskiplimit = \normallineskiplimit
      \multiply\lineskiplimit by #1 \divide\lineskiplimit by #2
      \parskip = \normalparskip
      \multiply\parskip by #1 \divide\parskip by #2
      \abovedisplayskip = \normaldisplayskip
      \multiply\abovedisplayskip by #1 \divide\abovedisplayskip by #2
      \belowdisplayskip = \abovedisplayskip
      \abovedisplayshortskip = \normaldispshortskip
      \multiply\abovedisplayshortskip by #1
        \divide\abovedisplayshortskip by #2
      \belowdisplayshortskip = \abovedisplayshortskip
      \advance\belowdisplayshortskip by \belowdisplayskip
      \divide\belowdisplayshortskip by 2
      \smallskipamount = \skipregister
      \multiply\smallskipamount by #1 \divide\smallskipamount by #2
      \medskipamount = \smallskipamount \multiply\medskipamount by 2
      \bigskipamount = \smallskipamount \multiply\bigskipamount by 4 }
\def\normalbaselines{ \baselineskip=\normalbaselineskip
   \lineskip=\normallineskip \lineskiplimit=\normallineskip
   \iftwelv@\else \multiply\baselineskip by 4 \divide\baselineskip by 5
     \multiply\lineskiplimit by 4 \divide\lineskiplimit by 5
     \multiply\lineskip by 4 \divide\lineskip by 5 \fi }
\Twelvepoint  
\interlinepenalty=50
\interfootnotelinepenalty=5000
\predisplaypenalty=9000
\postdisplaypenalty=500
\hfuzz=1pt
\vfuzz=0.2pt
\newdimen\HOFFSET  \HOFFSET=0pt
\newdimen\VOFFSET  \VOFFSET=0pt
\newdimen\HSWING   \HSWING=0pt
\dimen\footins=8in
%
%
%
\newskip\pagebottomfiller
\pagebottomfiller=\z@ plus \z@ minus \z@
\def\pagecontents{
   \ifvoid\topins\else\unvbox\topins\vskip\skip\topins\fi
   \dimen@ = \dp255 \unvbox255
   \vskip\pagebottomfiller
   \ifvoid\footins\else\vskip\skip\footins\footrule\unvbox\footins\fi
   \ifr@ggedbottom \kern-\dimen@ \vfil \fi }
\def\makeheadline{\vbox to 0pt{ \skip@=\topskip
      \advance\skip@ by -12pt \advance\skip@ by -2\normalbaselineskip
      \vskip\skip@ \line{\vbox to 12pt{}\the\headline} \vss
      }\nointerlineskip}
\def\makefootline{\baselineskip = 1.5\normalbaselineskip
                 \line{\the\footline}}
\newif\iffrontpage
\newif\ifp@genum
\def\nopagenumbers{\p@genumfalse}
\def\pagenumbers{\p@genumtrue}
\pagenumbers
\newtoks\paperheadline
\newtoks\paperfootline
\newtoks\letterheadline
\newtoks\letterfootline
\newtoks\letterinfo
\newtoks\date
\paperheadline={\hfil}
\paperfootline={\hss\iffrontpage\else\ifp@genum\tenrm\folio\hss\fi\fi}
\letterheadline{\iffrontpage \hfil \else
    \rm \ifp@genum page~~\folio\fi \hfil\the\date \fi}
\letterfootline={\iffrontpage\the\letterinfo\else\hfil\fi}
\letterinfo={\hfil}
\def\monthname{\rel@x\ifcase\month 0/\or January\or February\or
   March\or April\or May\or June\or July\or August\or September\or
   October\or November\or December\else\number\month/\fi}
\def\today{\monthname~\number\day, \number\year}
\date={\today}
\headline=\paperheadline 
\footline=\paperfootline 
\countdef\pageno=1      \countdef\pagen@=0
\countdef\pagenumber=1  \pagenumber=1
\def\advancepageno{\gl@bal\advance\pagen@ by 1
   \ifnum\pagenumber<0 \gl@bal\advance\pagenumber by -1
    \else\gl@bal\advance\pagenumber by 1 \fi
    \gl@bal\frontpagefalse  \swing@ }
\def\folio{\ifnum\pagenumber<0 \romannumeral-\pagenumber
           \else \number\pagenumber \fi }
\def\swing@{\ifodd\pagenumber \gl@bal\advance\hoffset by -\HSWING
             \else \gl@bal\advance\hoffset by \HSWING \fi }
\def\footrule{\dimen@=\prevdepth\nointerlineskip
   \vbox to 0pt{\vskip -0.25\baselineskip \hrule width 0.35\hsize \vss}
   \prevdepth=\dimen@ }
\let\footnotespecial=\rel@x
\newdimen\footindent
\footindent=24pt
\def\Textindent#1{\noindent\llap{#1\enspace}\ignorespaces}
\def\Vfootnote#1{\insert\footins\bgroup
   \interlinepenalty=\interfootnotelinepenalty \floatingpenalty=20000
   \singl@true\doubl@false\Tenpoint
   \splittopskip=\ht\strutbox \boxmaxdepth=\dp\strutbox
   \leftskip=\footindent \rightskip=\z@skip
   \parindent=0.5\footindent \parfillskip=0pt plus 1fil
   \spaceskip=\z@skip \xspaceskip=\z@skip \footnotespecial
   \Textindent{#1}\footstrut\futurelet\next\fo@t}

\def\vfootnote#1{\Vfootnote{${#1}$}}
\def\footnote#1{\attach{#1}\vfootnote{#1}}

\def\foot{\attach\footsymbolgen\vfootnote{\footsymbol}}
\let\footsymbol=\star
\newcount\lastf@@t           \lastf@@t=-1
\newcount\footsymbolcount    \footsymbolcount=0
\newif\ifPhysRev
\def\footsymbolgen{\bumpfootsymbolcount \generatefootsymbol \footsymbol }
\def\bumpfootsymbolcount{\rel@x
   \iffrontpage \bumpfootsymbolpos \else \advance\lastf@@t by 1
     \ifPhysRev \bumpfootsymbolneg \else \bumpfootsymbolpos \fi \fi
   \gl@bal\lastf@@t=\pagen@ }
\def\bumpfootsymbolpos{\ifnum\footsymbolcount <0
                            \gl@bal\footsymbolcount =0 \fi
    \ifnum\lastf@@t<\pagen@ \gl@bal\footsymbolcount=0
     \else \gl@bal\advance\footsymbolcount by 1 \fi }
\def\bumpfootsymbolneg{\ifnum\footsymbolcount >0
             \gl@bal\footsymbolcount =0 \fi
         \gl@bal\advance\footsymbolcount by -1 }
\def\fd@f#1 {\xdef\footsymbol{\mathchar"#1 }}
\def\generatefootsymbol{\ifcase\footsymbolcount \fd@f 13F \or \fd@f 279
        \or \fd@f 27A \or \fd@f 278 \or \fd@f 27B \else
        \ifnum\footsymbolcount <0 \fd@f{023 \number-\footsymbolcount }
         \else \fd@f 203 {\loop \ifnum\footsymbolcount >5
                \fd@f{203 \footsymbol } \advance\footsymbolcount by -1
                \repeat }\fi \fi }

\def\nonfrenchspacing{\sfcode`\.=3001 \sfcode`\!=3000 \sfcode`\?=3000
        \sfcode`\:=2000 \sfcode`\;=1500 \sfcode`\,=1251 }
\nonfrenchspacing
\newdimen\d@twidth
{\setbox0=\hbox{s.} \gl@bal\d@twidth=\wd0 \setbox0=\hbox{s}
        \gl@bal\advance\d@twidth by -\wd0 }
\def\removehglue{\loop \unskip \ifdim\lastskip >\z@ \repeat }
\def\roll@ver#1{\removehglue \nobreak \count255 =\spacefactor \dimen@=\z@
        \ifnum\count255 =3001 \dimen@=\d@twidth \fi
        \ifnum\count255 =1251 \dimen@=\d@twidth \fi
    \iftwelv@ \kern-\dimen@ \else \kern-0.83\dimen@ \fi
   #1\spacefactor=\count255 }
\def\step@ver#1{\rel@x \ifmmode #1\else \ifhmode
        \roll@ver{${}#1$}\else {\setbox0=\hbox{${}#1$}}\fi\fi }
\def\attach#1{\step@ver{\strut^{\mkern 2mu #1} }}
%
%
%
\newcount\chapternumber      \chapternumber=0
\newcount\sectionnumber      \sectionnumber=0
\newcount\equanumber         \equanumber=0
\let\chapterlabel=\rel@x
\let\sectionlabel=\rel@x
\newtoks\chapterstyle        \chapterstyle={\Number}
\newtoks\sectionstyle        \sectionstyle={\Number}
\newskip\chapterskip         \chapterskip=\bigskipamount
\newskip\sectionskip         \sectionskip=\medskipamount
\newskip\headskip            \headskip=8pt plus 3pt minus 3pt
\newdimen\chapterminspace    \chapterminspace=15pc
\newdimen\sectionminspace    \sectionminspace=10pc
\newdimen\referenceminspace  \referenceminspace=20pc
\newif\ifcn@                 \cn@true
\newif\ifcn@@                \cn@@false
\def\numberedchapters{\cn@true}
\def\unnumberedchapters{\cn@false\sequentialequations}
\def\chapterreset{\gl@bal\advance\chapternumber by 1
   \ifnum\equanumber<0 \else\gl@bal\equanumber=0\fi
   \sectionnumber=0 \let\sectionlabel=\rel@x
   \ifcn@ \gl@bal\cn@@true {\pr@tect
       \xdef\chapterlabel{\the\chapterstyle{\the\chapternumber}}}%
    \else \gl@bal\cn@@false \gdef\chapterlabel{\rel@x}\fi }
\def\@alpha#1{\count255='140 \advance\count255 by #1\char\count255}
 \def\alphabetic{\n@expand\@alpha}
\def\@Alpha#1{\count255='100 \advance\count255 by #1\char\count255}
 \def\Alphabetic{\n@expand\@Alpha}
\def\@Roman#1{\uppercase\expandafter{\romannumeral #1}}
 \def\Roman{\n@expand\@Roman}
\def\@roman#1{\romannumeral #1}    \def\roman{\n@expand\@roman}
\def\@number#1{\number #1}         \def\Number{\n@expand\@number}
\def\BLANK#1{\rel@x}               
\def\titleparagraphs{\interlinepenalty=9999
     \leftskip=0.03\hsize plus 0.22\hsize minus 0.03\hsize
     \rightskip=\leftskip \parfillskip=0pt
     \hyphenpenalty=9000 \exhyphenpenalty=9000
     \tolerance=9999 \pretolerance=9000
     \spaceskip=0.333em \xspaceskip=0.5em }
\def\titlestyle#1{\par\begingroup \titleparagraphs
     \iftwelv@\fourteenpoint\else\twelvepoint\fi
   \noindent #1\par\endgroup }
\def\spacecheck#1{\dimen@=\pagegoal\advance\dimen@ by -\pagetotal
   \ifdim\dimen@<#1 \ifdim\dimen@>0pt \vfil\break \fi\fi}
\def\chapter#1{\par \penalty-300 \vskip\chapterskip
   \spacecheck\chapterminspace
   \chapterreset \titlestyle{\ifcn@@\chapterlabel.~\fi #1}
   \nobreak\vskip\headskip \penalty 30000
   {\pr@tect\wlog{\string\chapter\space \chapterlabel}} }

\def\section#1{\par \ifnum\lastpenalty=30000\else
   \penalty-200\vskip\sectionskip \spacecheck\sectionminspace\fi
   \gl@bal\advance\sectionnumber by 1
   {\pr@tect
   \xdef\sectionlabel{\ifcn@@ \chapterlabel.\fi
       \the\sectionstyle{\the\sectionnumber}}%
   \wlog{\string\section\space \sectionlabel}}%
   \noindent {\caps\enspace\sectionlabel.~~#1}\par
   \nobreak\vskip\headskip \penalty 30000 }
\def\subsection#1{\par
   \ifnum\the\lastpenalty=30000\else \penalty-100\smallskip \fi
   \noindent\undertext{#1}\enspace \vadjust{\penalty5000}}

\def\undertext#1{\vtop{\hbox{#1}\kern 1pt \hrule}}
\def\APPENDIX#1#2{\par\penalty-300\vskip\chapterskip
   \spacecheck\chapterminspace \chapterreset \xdef\chapterlabel{#1}
   \titlestyle{APPENDIX #2} \nobreak\vskip\headskip \penalty 30000
   \wlog{\string\Appendix~\chapterlabel} }
\def\Appendix#1{\APPENDIX{#1}{#1}}
\def\appendix{\APPENDIX{A}{}}
%
%
%
\def\eqname#1{\rel@x {\pr@tect
  \ifnum\equanumber<0 \xdef#1{{\rm(\number-\equanumber)}}%
     \gl@bal\advance\equanumber by -1
  \else \gl@bal\advance\equanumber by 1
   \xdef#1{{\rm(\ifcn@@ \chapterlabel.\fi \number\equanumber)}}\fi
  }#1}
\def\eqinsert#1{\noalign{\dimen@=\prevdepth \nointerlineskip
   \setbox0=\hbox to\displaywidth{\hfil #1}
   \vbox to 0pt{\kern 0.5\baselineskip\hbox{$\!\box0\!$}\vss}
   \prevdepth=\dimen@}}
\def\leqinsert#1{\eqinsert{#1\hfill}}

%
%
\def\GENITEM#1;#2{\par \hangafter=0 \hangindent=#1
    \Textindent{$ #2 $}\ignorespaces}
\outer\def\newitem#1=#2;{\gdef#1{\GENITEM #2;}}

\newdimen\itemsize                \itemsize=30pt
\newitem\item=1\itemsize;
\newitem\sitem=1.75\itemsize;     
\newitem\ssitem=2.5\itemsize;     
\outer\def\newlist#1=#2&#3&#4;{\toks0={#2}\toks1={#3}%
   \count255=\escapechar \escapechar=-1
   \alloc@0\list\countdef\insc@unt\listcount     \listcount=0
   \edef#1{\par
      \countdef\listcount=\the\allocationnumber
      \advance\listcount by 1
      \hangafter=0 \hangindent=#4
      \Textindent{\the\toks0{\listcount}\the\toks1}}
   \expandafter\expandafter\expandafter
    \edef\c@t#1{begin}{\par
      \countdef\listcount=\the\allocationnumber \listcount=1
      \hangafter=0 \hangindent=#4
      \Textindent{\the\toks0{\listcount}\the\toks1}}
   \expandafter\expandafter\expandafter
    \edef\c@t#1{con}{\par \hangafter=0 \hangindent=#4 \noindent}
   \escapechar=\count255}
\def\c@t#1#2{\csname\string#1#2\endcsname}
\newlist\point=\Number&.&1.0\itemsize;
\newlist\subpoint=(\alphabetic&)&1.75\itemsize;
\newlist\subsubpoint=(\roman&)&2.5\itemsize;
%

%
%
%
%
\newcount\referencecount     \referencecount=0
\newcount\lastrefsbegincount \lastrefsbegincount=0
\newif\ifreferenceopen       \newwrite\referencewrite
\newdimen\refindent          \refindent=30pt
\def\normalrefmark#1{\attach{\scriptscriptstyle [ #1 ] }}
\let\PRrefmark=\attach
\def\NPrefmark#1{\step@ver{{\;[#1]}}}
\def\refmark#1{\rel@x\ifPhysRev\PRrefmark{#1}\else\normalrefmark{#1}\fi}
\def\refend@{\refmark{\number\referencecount}}
\def\refend{\refend@{}\space }
\def\refsend{\refmark{\count255=\referencecount
   \advance\count255 by-\lastrefsbegincount
   \ifcase\count255 \number\referencecount
   \or \number\lastrefsbegincount,\number\referencecount
   \else \number\lastrefsbegincount-\number\referencecount \fi}\space }
\def\REFNUM#1{\rel@x \gl@bal\advance\referencecount by 1
    \xdef#1{\the\referencecount }}
\def\Refnum#1{\REFNUM #1\refend@ } 
\def\REF#1{\REFNUM #1\R@FWRITE\ignorespaces}
\def\Ref#1{\Refnum #1\REFWRITE }
\def\ref{\Ref\?}
\def\REFS#1{\REFNUM #1\gl@bal\lastrefsbegincount=\referencecount
    \REFWRITE }

\def\r@fitem#1{\par \hangafter=0 \hangindent=\refindent \Textindent{#1}}
\def\refitem#1{\r@fitem{#1.}}
\def\NPrefitem#1{\r@fitem{[#1]}}
\def\NPrefs{\let\refmark=\NPrefmark \let\refitem=NPrefitem}
\def\REFWRITE{\R@FWRITE\rel@x }
\def\R@FWRITE#1{\ifreferenceopen \else \gl@bal\referenceopentrue
     \immediate\openout\referencewrite=\jobname.refs
     \toks@={\begingroup \refoutspecials \catcode`\^^M=10 }%
     \immediate\write\referencewrite{\the\toks@}\fi
    \immediate\write\referencewrite{\noexpand\refitem %
                                    {\the\referencecount}}%
    \p@rse@ndwrite \referencewrite #1}
\begingroup
 \catcode`\^^M=\active \let^^M=\relax %
 \gdef\p@rse@ndwrite#1#2{\begingroup \catcode`\^^M=12 \newlinechar=`\^^M%
         \chardef\rw@write=#1\sc@nlines#2}%
 \gdef\sc@nlines#1#2{\sc@n@line \g@rbage #2^^M\endsc@n \endgroup #1}%
 \gdef\sc@n@line#1^^M{\expandafter\toks@\expandafter{\deg@rbage #1}%
         \immediate\write\rw@write{\the\toks@}%
         \futurelet\n@xt \sc@ntest }%
\endgroup
\def\sc@ntest{\ifx\n@xt\endsc@n \let\n@xt=\rel@x
       \else \let\n@xt=\sc@n@notherline \fi \n@xt }
\def\sc@n@notherline{\sc@n@line \g@rbage }
\def\deg@rbage#1{}
\let\g@rbage=\relax    \let\endsc@n=\relax
\def\refout{\par\penalty-400\vskip\chapterskip
   \spacecheck\referenceminspace
   \ifreferenceopen \Closeout\referencewrite \referenceopenfalse \fi
   \line{\fourteenrm\hfil REFERENCES\hfil}\vskip\headskip
   \input \jobname.refs
   }
\def\refoutspecials{\sfcode`\.=1000 \interlinepenalty=1000
         \rightskip=\z@ plus 1em minus \z@ }
\def\Closeout#1{\toks0={\par\endgroup}\immediate\write#1{\the\toks0}%
   \immediate\closeout#1}
%
%
\newcount\figurecount     \figurecount=0
\newcount\tablecount      \tablecount=0
\newif\iffigureopen       \newwrite\figurewrite
\newif\iftableopen        \newwrite\tablewrite
\def\FIGNUM#1{\rel@x \gl@bal\advance\figurecount by 1
    \xdef#1{\the\figurecount}}
\def\FIGURE#1{\FIGNUM #1\F@GWRITE\ignorespaces }

\def\Fig{\FIGNUM\?Figure~\?\FIGWRITE }

\def\figitem#1{\r@fitem{#1)}}
\def\FIGWRITE{\F@GWRITE\rel@x }
\def\TABNUM#1{\rel@x \gl@bal\advance\tablecount by 1
    \xdef#1{\the\tablecount}}
\def\TABLE#1{\TABNUM #1\T@BWRITE\ignorespaces }
\def\Table{\TABNUM\?Table~\?\TABWRITE }
\def\tabitem#1{\r@fitem{#1:}}
\def\TABWRITE{\T@BWRITE\rel@x }
\def\F@GWRITE#1{\iffigureopen \else \gl@bal\figureopentrue
     \immediate\openout\figurewrite=\jobname.figs
     \toks@={\begingroup \catcode`\^^M=10 }%
     \immediate\write\figurewrite{\the\toks@}\fi
    \immediate\write\figurewrite{\noexpand\figitem %
                                 {\the\figurecount}}%
    \p@rse@ndwrite \figurewrite #1}
\def\T@BWRITE#1{\iftableopen \else \gl@bal\tableopentrue
     \immediate\openout\tablewrite=\jobname.tabs
     \toks@={\begingroup \catcode`\^^M=10 }%
     \immediate\write\tablewrite{\the\toks@}\fi
    \immediate\write\tablewrite{\noexpand\tabitem %
                                 {\the\tablecount}}%
    \p@rse@ndwrite \tablewrite #1}
\def\figout{\par\penalty-400
   \vskip\chapterskip\spacecheck\referenceminspace
   \iffigureopen \Closeout\figurewrite \figureopenfalse \fi
   \line{\fourteenrm\hfil FIGURE CAPTIONS\hfil}\vskip\headskip
   \input \jobname.figs
   }
\def\tabout{\par\penalty-400
   \vskip\chapterskip\spacecheck\referenceminspace
   \iftableopen \Closeout\tablewrite \tableopenfalse \fi
   \line{\fourteenrm\hfil TABLE CAPTIONS\hfil}\vskip\headskip
   \input \jobname.tabs
   }
%
%
%
\newbox\picturebox
\def\p@cht{\ht\picturebox }
\def\p@cwd{\wd\picturebox }
\def\p@cdp{\dp\picturebox }
\newdimen\xshift
\newdimen\yshift
\newdimen\captionwidth
\newskip\captionskip
\captionskip=15pt plus 5pt minus 3pt
\def\fullwidth{\captionwidth=\hsize }
\newtoks\Caption
\newif\ifcaptioned
\newif\ifselfcaptioned
\def\caption{\captionedtrue \Caption }
\newcount\linesabove
\newif\iffileexists
\newtoks\picfilename
\def\fil@#1 {\fileexiststrue \picfilename={#1}}
\def\file#1{\if=#1\let\n@xt=\fil@ \else \def\n@xt{\fil@ #1}\fi \n@xt }
\def\pl@t{\begingroup \pr@tect
    \setbox\picturebox=\hbox{}\fileexistsfalse
    \let\height=\p@cht \let\width=\p@cwd \let\depth=\p@cdp
    \xshift=\z@ \yshift=\z@ \captionwidth=\z@
    \Caption={}\captionedfalse
    \linesabove =0 \picturedefault }
\def\plot{\pl@t \selfcaptionedfalse }
\def\Picture#1{\gl@bal\advance\figurecount by 1
    \xdef#1{\the\figurecount}\pl@t \selfcaptionedtrue }

\def\s@vepicture{\iffileexists \parsefilename \redopicturebox \fi
   \ifdim\captionwidth>\z@ \else \captionwidth=\p@cwd \fi
   \xdef\lastpicture{\iffileexists
        \setbox0=\hbox{\raise\the\yshift \vbox{%
              \moveright\the\xshift\hbox{\picturedefinition}}}%
        \else \setbox0=\hbox{}\fi
         \ht0=\the\p@cht \wd0=\the\p@cwd \dp0=\the\p@cdp
         \vbox{\hsize=\the\captionwidth \line{\hss\box0 \hss }%
              \ifcaptioned \vskip\the\captionskip \noexpand\Tenpoint
                \ifselfcaptioned Figure~\the\figurecount.\enspace \fi
                \the\Caption \fi }}%
    \endgroup }
\let\endpicture=\s@vepicture
\def\savepicture#1{\s@vepicture \global\let#1=\lastpicture }
\def\displaypicture{\fullwidth \s@vepicture $$\lastpicture $${}}
\def\toppicture{\fullwidth \s@vepicture \topinsert
    \lastpicture \medskip \endinsert }
\def\midpicture{\fullwidth \s@vepicture \midinsert
    \lastpicture \endinsert }
%
%
\def\leftpicture{\pres@tpicture
    \dimen@i=\hsize \advance\dimen@i by -\dimen@ii
    \setbox\picturebox=\hbox to \hsize {\box0 \hss }%
    \wr@paround }
\def\rightpicture{\pres@tpicture
    \dimen@i=\z@
    \setbox\picturebox=\hbox to \hsize {\hss \box0 }%
    \wr@paround }
\def\pres@tpicture{\gl@bal\linesabove=\linesabove
    \s@vepicture \setbox\picturebox=\vbox{
         \kern \linesabove\baselineskip \kern 0.3\baselineskip
         \lastpicture \kern 0.3\baselineskip }%
    \dimen@=\p@cht \dimen@i=\dimen@
    \advance\dimen@i by \pagetotal
    \par \ifdim\dimen@i>\pagegoal \vfil\break \fi
    \dimen@ii=\hsize
    \advance\dimen@ii by -\parindent \advance\dimen@ii by -\p@cwd
    \setbox0=\vbox to\z@{\kern-\baselineskip \unvbox\picturebox \vss }}
\def\wr@paround{\Caption={}\count255=1
    \loop \ifnum \linesabove >0
         \advance\linesabove by -1 \advance\count255 by 1
         \advance\dimen@ by -\baselineskip
         \expandafter\Caption \expandafter{\the\Caption \z@ \hsize }%
      \repeat
    \loop \ifdim \dimen@ >\z@
         \advance\count255 by 1 \advance\dimen@ by -\baselineskip
         \expandafter\Caption \expandafter{%
             \the\Caption \dimen@i \dimen@ii }%
      \repeat
    \edef\n@xt{\parshape=\the\count255 \the\Caption \z@ \hsize }%
    \par\noindent \n@xt \strut \vadjust{\box\picturebox }}
\let\picturedefault=\relax
\let\parsefilename=\relax
\def\redopicturebox{\let\picturedefinition=\rel@x
   \errhelp=\disabledpictures
   \errmessage{This version of TeX cannot handle pictures.  Sorry.}}
\newhelp\disabledpictures
     {You will get a blank box in place of your picture.}
%
%
%
%
%
%
%
%
%
%
\def\FRONTPAGE{\ifvoid255\else\vfill\penalty-20000\fi
   \gl@bal\pagenumber=1     \gl@bal\chapternumber=0
   \gl@bal\equanumber=0     \gl@bal\sectionnumber=0
   \gl@bal\referencecount=0 \gl@bal\figurecount=0
   \gl@bal\tablecount=0     \gl@bal\frontpagetrue
   \gl@bal\lastf@@t=0       \gl@bal\footsymbolcount=0
   \gl@bal\cn@@false }

\def\papers{\papersize\headline=\paperheadline\footline=\paperfootline}
\def\papersize{\hsize=35pc \vsize=50pc \hoffset=0pc \voffset=1pc
   \advance\hoffset by\HOFFSET \advance\voffset by\VOFFSET
   \pagebottomfiller=0pc
   \skip\footins=\bigskipamount \normalspace }
\papers  
%
%
\newskip\lettertopskip       \lettertopskip=20pt plus 50pt
\newskip\letterbottomskip    \letterbottomskip=\z@ plus 100pt
\newskip\signatureskip       \signatureskip=40pt plus 3pt
\def\lettersize{\hsize=6.5in \vsize=8.5in \hoffset=0in \voffset=0.5in
   \advance\hoffset by\HOFFSET \advance\voffset by\VOFFSET
   \pagebottomfiller=\letterbottomskip
   \skip\footins=\smallskipamount \multiply\skip\footins by 3
   \singlespace }
\def\MEMO{\lettersize \headline=\letterheadline \footline={\hfil }%
   \let\rule=\memorule \FRONTPAGE \memohead }

\def\memodate{\afterassignment\MEMO \date }
\def\memit@m#1{\smallskip \hangafter=0 \hangindent=1in
    \Textindent{\caps #1}}
\def\subject{\memit@m{Subject:}}
\def\topic{\memit@m{Topic:}}
\def\from{\memit@m{From:}}
\def\to{\rel@x \ifmmode \rightarrow \else \memit@m{To:}\fi }
\def\memorule{\medskip\hrule height 1pt\bigskip}  
\def\memohead{\centerline{\fourteenrm MEMORANDUM}}
\newwrite\labelswrite
\newtoks\rw@toks
\def\letters{\lettersize
   \headline=\letterheadline \footline=\letterfootline
   \immediate\openout\labelswrite=\jobname.lab}

\let\letterhead=\rel@x
\def\addressee#1{\medskip\line{\hskip 0.75\hsize plus\z@ minus 0.25\hsize
                               \the\date \hfil }%
   \vskip \lettertopskip
   \ialign to\hsize{\strut ##\hfil\tabskip 0pt plus \hsize \crcr #1\crcr}
   \writelabel{#1}\medskip \noindent\hskip -\spaceskip \ignorespaces }
\def\rwl@begin#1\cr{\rw@toks={#1\crcr}\rel@x
   \immediate\write\labelswrite{\the\rw@toks}\futurelet\n@xt\rwl@next}
\def\rwl@next{\ifx\n@xt\rwl@end \let\n@xt=\rel@x
      \else \let\n@xt=\rwl@begin \fi \n@xt}
\let\rwl@end=\rel@x
\def\writelabel#1{\immediate\write\labelswrite{\noexpand\labelbegin}
     \rwl@begin #1\cr\rwl@end
     \immediate\write\labelswrite{\noexpand\labelend}}
\newtoks\FromAddress         \FromAddress={}
\newtoks\sendername          \sendername={}
\newbox\FromLabelBox
\newdimen\labelwidth          \labelwidth=6in
\def\makelabels{\afterassignment\Makelabels \sendername=}
\def\Makelabels{\FRONTPAGE \letterinfo={\hfil } \MakeFromBox
     \immediate\closeout\labelswrite  \input \jobname.lab\vfil\eject}
\let\labelend=\rel@x
\def\labelbegin#1\labelend{\setbox0=\vbox{\ialign{##\hfil\cr #1\crcr}}
     \MakeALabel }
\def\MakeFromBox{\gl@bal\setbox\FromLabelBox=\vbox{\Tenpoint
     \ialign{##\hfil\cr \the\sendername \the\FromAddress \crcr }}}
\def\MakeALabel{\vskip 1pt \hbox{\vrule \vbox{
        \hsize=\labelwidth \hrule\bigskip
        \leftline{\hskip 1\parindent \copy\FromLabelBox}\bigskip
        \centerline{\hfil \box0 } \bigskip \hrule
        }\vrule } \vskip 1pt plus 1fil }
\def\signed#1{\par \nobreak \bigskip \dt@pfalse \begingroup
  \everycr={\noalign{\nobreak
            \ifdt@p\vskip\signatureskip\gl@bal\dt@pfalse\fi }}%
  \tabskip=0.5\hsize plus \z@ minus 0.5\hsize
  \halign to\hsize {\strut ##\hfil\tabskip=\z@ plus 1fil minus \z@\crcr
          \noalign{\gl@bal\dt@ptrue}#1\crcr }%
  \endgroup \bigskip }
\newbox\letterb@x
\def\lettertext{\par \vskip\parskip \unvcopy\letterb@x \par }
\def\multiletter{\setbox\letterb@x=\vbox\bgroup
      \everypar{\vrule height 1\baselineskip depth 0pt width 0pt }
      \singlespace \topskip=\baselineskip }
\def\letterend{\par\egroup}
%
%
%
\newskip\frontpageskip
\newtoks\Pubnum   
\newtoks\Pubtype  \let\pubtype=\Pubtype
\newif\ifp@bblock  \p@bblocktrue
\def\PH@SR@V{\doubl@true \baselineskip=24.1pt plus 0.2pt minus 0.1pt
             \parskip= 3pt plus 2pt minus 1pt }
\def\PHYSREV{\papers\PhysRevtrue\PH@SR@V}

\def\titlepage{\FRONTPAGE\papers\ifPhysRev\PH@SR@V\fi
   \ifp@bblock\p@bblock \else\hrule height\z@ \rel@x \fi }
\def\nopubblock{\p@bblockfalse}
\def\endpage{\vfil\break}
\frontpageskip=12pt plus .5fil minus 2pt
\Pubtype={}
\Pubnum={}
\def\p@bblock{\begingroup \tabskip=\hsize minus \hsize
   \baselineskip=1.5\ht\strutbox \topspace-2\baselineskip
   \halign to\hsize{\strut ##\hfil\tabskip=0pt\crcr
       \the\Pubnum\crcr\the\date\crcr\the\pubtype\crcr}\endgroup}
\def\title#1{\vskip\frontpageskip \titlestyle{#1} \vskip\headskip }
\def\author#1{\vskip\frontpageskip\titlestyle{\twelvecp #1}\nobreak}

\def\address#1{\par\kern 5pt\titlestyle{\twelvepoint\it #1}}
\def\andaddress{\par\kern 5pt \centerline{\sl and} \address}

\def\abstract{\par\dimen@=\prevdepth \hrule height\z@ \prevdepth=\dimen@
   \vskip\frontpageskip\centerline{\fourteenrm ABSTRACT}\vskip\headskip }

%
%
%
\def\ie{\hbox{\it i.e.}}

\def\\{\rel@x \ifmmode \backslash \else {\tt\char`\\}\fi }
\def\sequentialequations{\rel@x \if\equanumber<0 \else
  \gl@bal\equanumber=-\equanumber \gl@bal\advance\equanumber by -1 \fi }
\def\journal#1&#2(#3){\begingroup \let\journal=\dummyj@urnal
    \unskip, \sl #1\unskip~\bf\ignorespaces #2\rm
    (\afterassignment\j@ur \count255=#3), \endgroup\ignorespaces }
\def\j@ur{\ifnum\count255<100 \advance\count255 by 1900 \fi
          \number\count255 }
\def\dummyj@urnal{%
    \toks@={Reference foul up: nested \journal macros}%
    \errhelp={Your forgot & or ( ) after the last \journal}%
    \errmessage{\the\toks@ }}

\def\topspace{\hrule height 0pt depth 0pt \vskip}

\def\Buildrel#1\under#2{\mathrel{\mathop{#2}\limits_{#1}}}
\def\becomes#1{\mathchoice{\becomes@\scriptstyle{#1}}
   {\becomes@\scriptstyle{#1}} {\becomes@\scriptscriptstyle{#1}}
   {\becomes@\scriptscriptstyle{#1}}}
\def\becomes@#1#2{\mathrel{\setbox0=\hbox{$\m@th #1{\,#2\,}$}%
        \mathop{\hbox to \wd0 {\rightarrowfill}}\limits_{#2}}}

\let\int=\intop         
\def\lsim{\mathrel{\mathpalette\@versim<}}
\def\gsim{\mathrel{\mathpalette\@versim>}}
\def\@versim#1#2{\vcenter{\offinterlineskip
        \ialign{$\m@th#1\hfil##\hfil$\crcr#2\crcr\sim\crcr } }}
\def\big#1{{\hbox{$\left#1\vbox to 0.85\b@gheight{}\right.\n@space$}}}
\def\Big#1{{\hbox{$\left#1\vbox to 1.15\b@gheight{}\right.\n@space$}}}
\def\bigg#1{{\hbox{$\left#1\vbox to 1.45\b@gheight{}\right.\n@space$}}}
\def\Bigg#1{{\hbox{$\left#1\vbox to 1.75\b@gheight{}\right.\n@space$}}}
\def\){\mskip 2mu\nobreak }
%
%
%
\let\sec@nt=\sec
\def\sec{\rel@x\ifmmode\let\n@xt=\sec@nt\else\let\n@xt\section\fi\n@xt}
\def\obsolete#1{\message{Macro \string #1 is obsolete.}}
\def\firstsec#1{\obsolete\firstsec \section{#1}}
\def\firstsubsec#1{\obsolete\firstsubsec \subsection{#1}}
\def\thispage#1{\obsolete\thispage \gl@bal\pagenumber=#1\frontpagefalse}
\def\thischapter#1{\obsolete\thischapter \gl@bal\chapternumber=#1}
\def\splitout{\obsolete\splitout\rel@x}
\def\prop{\obsolete\prop \propto }
\def\nextequation#1{\obsolete\nextequation \gl@bal\equanumber=#1
   \ifnum\the\equanumber>0 \gl@bal\advance\equanumber by 1 \fi}
\def\BOXITEM{\afterassigment\B@XITEM\setbox0=}
\def\B@XITEM{\par\hangindent\wd0 \noindent\box0 }
%
%
%
\def\phyzzx{PHY\setbox0=\hbox{Z}\copy0 \kern-0.5\wd0 \box0 X}
        
\everyjob{\xdef\today{\monthname~\number\day, \number\year}
 }
\message{ by V.K.}
%
%
\newcount	 \ObjClass
\chardef\ClassNum	= 0
\chardef\ClassMisc	= 1
\chardef\ClassEqn	= 2
\chardef\ClassRef	= 3
\chardef\ClassFig	= 4
\chardef\ClassTbl	= 5
\chardef\ClassThm	= 6
\chardef\ClassStyle     = 7
\chardef\ClassDef       = 8
\edef\NumObj	{\ObjClass = \ClassNum   \relax}
\edef\MiscObj	{\ObjClass = \ClassMisc  \relax}
\edef\EqnObj	{\ObjClass = \ClassEqn   \relax}
\edef\RefObj	{\ObjClass = \ClassRef   \relax}
\edef\FigObj	{\ObjClass = \ClassFig   \relax}
\edef\TblObj	{\ObjClass = \ClassTbl   \relax}
\edef\ThmObj	{\ObjClass = \ClassThm   \relax}
\edef\StyleObj  {\ObjClass = \ClassStyle \relax}
\edef\DefObj    {\ObjClass = \ClassDef   \relax}
%
%
\def\gobble	 #1{}%
\def\trimspace   #1 \end{#1}%
\def\ifundefined #1{\expandafter \ifx \csname#1\endcsname \relax}%
\def\trimprefix  #1_#2\end{\expandafter \string \csname #2\endcsname}%
\def\skipspace #1#2#3\end%
    {%
    \def \temp {#2}%
    \ifx \temp\space \skipspace #1#3\end
    \else \gdef #1{#2#3}\fi
    }%
\def\stylename#1{\expandafter\expandafter\expandafter
    \gobble\expandafter\string\the#1}
\ifundefined {protect} \let\protect=\relax \fi
\catcode`\@=11
\let\rel@x=\relax
\def\relaxtest{\rel@x}
\catcode`\@=12
\def\checkchapterlabel%
    {%
        {\protect\if\chapterlabel\relaxtest
	\global\let\chapterlabel=\relax\fi}
    }%
\begingroup
\catcode`\<=1 \catcode`\{=12
\catcode`\>=2 \catcode`\}=12
\xdef\LBrace<{>%
\xdef\RBrace<}>%
\endgroup
%
%
\newcount\equanumber \equanumber=0
\newcount\eqnumber   \eqnumber=0
\newif\ifleftnumbers \leftnumbersfalse

\def\(#1)%
     {%
        \ifnum \equanumber<0 \eqnumber=-\equanumber
	    \advance\eqnumber by -1 \else
            \eqnumber = \equanumber\fi
        \ifmmode\ifinner(\eqnum {#1})\else
        \ifleftnumbers\leqno(\eqnum {#1})\ifdraft{\rm[#1]}\fi
            \else\eqno(\eqnum {#1})\ifdraft{\rm[#1]}\fi\fi\fi
	\else(\eqnum {#1})\fi\ifnum%
	    \equanumber<0 \global\equanumber=-\eqnumber\global\advance
            \equanumber by -1\else\global\equanumber=\eqnumber\fi
     }%
\def\mideq(#1)%
     {%
	\ifleftnumbers \leqinsert{$\(#1)$} \else
	\eqinsert{$\(#1)$} \fi
     }%
\def\eqnum #1%
    {%
    \LookUp Eq_#1 \using\eqnumber\neweqnum
    {\rm \label}%
    }%
\def\neweqnum #1#2%
    {%
    \checkchapterlabel
    {\protect\xdef\eqnoprefix{\ifundefined{chapterlabel}
	\else\chapterlabel.\fi}}
    \ifmmode \xdef #1{\eqnoprefix #1}
        \else\message{Undefined equation \string#1 in non-math mode.}%
	     \xdef #1{\relax}
	     \global\advance \eqnumber by -1
        \fi
    \EqnObj \SaveObject{#1}{#2}
    }%
\everydisplay = {\expandafter \let\csname Eq_\endcsname=\relax
		 \expandafter \let\csname Eq_?\endcsname=\relax}%
%
%
\newcount\tablecount \tablecount=0
\def\Table  #1{Table~\tblnum {#1}}%
\def\tblnum #1{\TblObj \LookUp Tbl_#1 \using\tablecount
	\SaveObject \label\ifdraft [#1]\fi}%
\def\tbldef #1{\TblObj \SaveContents {Tbl_#1}}%
%
%
\def\inserttable #1#2#3%
    {%
    \tbldef {#1}{#3}\goodbreak%
    \midinsert
	\smallskip
	\hbox{\singlespace
	      \vtop{\titlestyle{{\Tenpoint{\caps\Table{#1}}\break #3}}}
    	     }%
	#2
    	\smallskip%
    \endinsert
    }%
\def\topinserttable #1#2#3%
    {%
    \tbldef {#1}{#3}\goodbreak%
    \topinsert
	\smallskip
	\hbox{\singlespace
	      \vtop{\titlestyle{{\Tenpoint{\caps\Table{#1}}\break #3}}}
    	     }%
	#2
    	\smallskip%
    \endinsert
    }%
%
%
\newcount\figurecount \figurecount=0
\def\Fig    #1{Fig.~\fignum {#1}}%
\def\fignum #1{\FigObj \LookUp Fig_#1 \using\figurecount
     \SaveObject \label\ifdraft [#1]\fi}%
\def\figdef #1{\FigObj \SaveContents {Fig_#1}}%
%
%
\def\insertfigure #1#2#3%
    {%
    \figdef {#1}{#3}%
    \midinsert
	\bigskip
	#2
	\hbox{	\singlespace
		\hskip 0.4in
		\vtop{\parshape=2 0pt 362pt 32pt 330pt
		      \noindent{\Tenpoint{\caps\Fig{#1}}.\enspace #3}}
		\hfil}
    	\smallskip%
    \endinsert
    }%
\def\topinsertfigure #1#2#3%
    {%
    \figdef {#1}{#3}%
    \topinsert
	\bigskip
	#2
	\hbox{	\singlespace
		\hskip 0.4in
		\vtop{\parshape=2 0pt 362pt 32pt 330pt
		      \noindent{\Tenpoint{\caps\Fig{#1}}.\enspace #3}}
		\hfil}
    	\smallskip%
    \endinsert
    }%
%
%
%
\newcount\theoremcount \theoremcount=0
\def\prop		#1{prop.~\thmnum {#1}}%
\def\thmnum #1{\ThmObj \LookUp Thm_#1 \using\theoremcount
     \SaveObject \label\ifdraft [#1]\fi}%
%
%
\newcount\referencecount \referencecount=0
\newcount\refsequence	\refsequence=0
\newcount\lastrefno	\lastrefno=-1
%
\def\NPrefs{\let\refmark=\NPrefmark \let\refitem=\NPrefitem}
\def\normalrefmark#1{\attach{[ #1 ]}}
\def\refsymbol#1{\refrange#1-\end}%
\def\[#1]#2%
	{%
	\if.#2\rlap.\refmark{\refsymbol{#1}}\let\refendtok=\relax%
	\else\if,#2\rlap,\refmark{\refsymbol{#1}}\let\refendtok=\relax%
    	\else\refmark{\refsymbol{#1}}\let\refendtok=#2\fi\fi%
	\discretionary{}{}{}\refendtok}%
\def\refrange #1-#2\end%
    {%
    \refnums #1,\end
    \def \temp {#2}%
    \ifx \temp\empty \else -\expandafter\refrange \temp\end \fi
    }%
\def\refnums #1,#2\end%
    {%
    \def \temp {#1}%
    \ifx \temp\empty \else \skipspace \temp#1\end\fi
    \ifx \temp\empty
	\ifcase \refsequence
	    \or\or ,\number\lastrefno
	    \else  -\number\lastrefno
	\fi
	\global\lastrefno = -1
	\global\refsequence = 0
    \else
	\RefObj \edef\temp {Ref_\temp\space}%
	\expandafter \LookUp \temp \using\referencecount\SaveObject
	\global\advance \lastrefno by 1
	\edef \temp {\number\lastrefno}%
	\ifx \label\temp
	    \global\advance\refsequence by 1
	\else
	    \global\advance\lastrefno by -1
	    \ifcase \refsequence
		\or ,%
		\or ,\number\lastrefno,%
	    \else   -\number\lastrefno,%
	    \fi
	    \label
	    \global\refsequence = 1
	    \ifx\suffix\empty
		\global\lastrefno = \label
	    \else
		\global\lastrefno = -1
	    \fi
	\fi
	\refnums #2,\end
    \fi
    }%
%
%
%
%
\def\refdef #1{\RefObj \SaveContents {Ref_#1}}%
\def\reflist  {\RefObj \ListObjects}%
\def\Refer #1{Ref.~\refsymbol{#1}}%
%
%
\newif\ifSaveFile
\newif\ifnotskip
\newwrite\SaveFile
\let\IfSelect=\iftrue
\edef\savefilename {\jobname.aux}%
\def\Def#1#2%
    {%
    \expandafter\gdef\noexpand#1{#2}%
    \DefObj \SaveObject {#2}{\expandafter\gobble\string#1}%
}%
\def\savestate%
    {%
    \ifundefined {chapternumber} \else
	\NumObj \SaveObject {\number\chapternumber}{chapternumber} \fi
        \ifundefined {appendixnumber} \else
	\NumObj \SaveObject {\number\appendixnumber}{appendixnumber} \fi
    \ifundefined {sectionnumber} \else
	\NumObj \SaveObject {\number\sectionnumber}{sectionnumber} \fi
    \ifundefined {pagenumber} \else
	\advance\pagenumber by 1
	\NumObj \SaveObject {\number\pagenumber}{pagenumber}%
	\advance\pagenumber by -1 \fi
    \NumObj \SaveObject {\number\equanumber}{equanumber}%
    \NumObj \SaveObject {\number\tablecount}{tablecount}%
    \NumObj \SaveObject {\number\figurecount}{figurecount}%
    \NumObj \SaveObject {\number\theoremcount}{theoremcount}%
    \NumObj \SaveObject {\number\referencecount}{referencecount}%
    \checkchapterlabel
    \ifundefined {chapterlabel} \else
	{\protect\xdef\chaplabel{\chapterlabel}}
	\MiscObj \SaveObject \chaplabel {chapterlabel} \fi
    \ifundefined {chapterstyle} \else
   	\StyleObj \SaveObject {\stylename{\chapterstyle}}{chapterstyle} \fi
    \ifundefined {appendixstyle} \else
	\StyleObj \SaveObject {\stylename{\appendixstyle}}{appendixstyle}\fi
}%
\def\Contents #1{\ObjClass=-#1 \SaveContents}%
\def\Define #1#2#3%
    {%
    \ifnum #1=\ClassNum
	\global \csname#2\endcsname = #3 %
    \else \ifnum #1=\ClassStyle
	\global \csname#2\endcsname\expandafter=
	\expandafter{\csname#3\endcsname} %
    \else \ifnum #1=\ClassDef
        \expandafter\gdef\csname#2\endcsname{#3} %
    \else
	\expandafter\xdef \csname#2\endcsname {#3} \fi\fi\fi %
    \ObjClass=#1 \SaveObject {#3}{#2}%
    }%
\def\SaveObject #1#2%
    {%
    \ifSaveFile \else \OpenSaveFile \fi
    \immediate\write\SaveFile
	{%
	\noexpand\IfSelect\noexpand\Define
	{\the\ObjClass}{#2}{#1}\noexpand\fi
	}%
    }%
\def\SaveContents #1%
    {%
    \ifSaveFile \else \OpenSaveFile \fi
    \BreakLine
    \SaveLine {#1}%
    }%
\begingroup
    \catcode`\^^M=\active %
\gdef\BreakLine %
    {%
    \begingroup %
    \catcode`\^^M=\active %
    \newlinechar=`\^^M %
    }%
\gdef\SaveLine #1#2%
    {%
    \toks255={#2}%
    \immediate\write\SaveFile %
	{%
	\noexpand\IfSelect\noexpand\Contents
	{-\the\ObjClass}{#1}\LBrace\the\toks255\RBrace\noexpand\fi%
	}%
    \endgroup %
    }%
\endgroup
\def\ListObjects #1%
    {%
    \ifSaveFile \CloseSaveFile \fi
    \let \IfSelect=\GetContents \ReadFileList #1,\savefilename,\end
    \let \IfSelect=\IfDoObject  \input \savefilename
    \let \IfSelect=\iftrue
    }%
\def\ReadFileList #1,#2\end%
    {%
    \def \temp {#1}%
    \ifx \temp\empty \else \skipspace \temp#1\end \fi
    \ifx \temp\empty \else \input #1 \fi
    \def \temp {#2}%
    \ifx \temp\empty \else \ReadFileList #2\end \fi
    }%
\def\GetContents #1#2#3%
    {%
    \notskipfalse
    \ifnum \ObjClass=-#2
	\expandafter\ifx \csname #3\endcsname \relax \else \notskiptrue \fi
    \fi
    \ifnotskip \expandafter \DefContents \csname #3_\endcsname
    }%
\def\DefContents #1#2{\toks255={#2} \xdef #1{\the\toks255}}%
\def\IfDoObject #1#2%
    {%
    \notskipfalse \ifnum \ObjClass=#2 \notskiptrue\fi \ifnotskip \DoObject
    }%
\def\DoObject #1#2%
    {%
    \ifnum \ObjClass = \ClassTbl	\par\noindent Table~#2.
    \else \ifnum \ObjClass = \ClassFig	\par\noindent Figure~#2.
    \else \ifnum \ObjClass = \ClassRef  \refitem{#2}
    \else \item {#2.}
    \fi\fi\fi
    \ifdraft\edef\temp {\trimprefix #1\end}[\expandafter\gobble \temp]~\fi
    \expandafter\ifx \csname #1_\endcsname \relax
	\ifdraft\relax\else\edef\temp {\trimprefix #1\end}%
	[\expandafter\gobble \temp]\fi%
    \else
	\csname #1_\endcsname
    \fi
    }%
\def\OpenSaveFile   {\immediate\openout\SaveFile=\savefilename
		     \global\SaveFiletrue}%
\def\CloseSaveFile  {\immediate\closeout\SaveFile \global\SaveFilefalse}%
%
%
\def\LookUp #1 #2\using#3#4%
    {%
    \expandafter \ifx\csname#1\endcsname \relax
	\global\advance #3 by 1
	\expandafter \xdef \csname#1\endcsname {\number #3}%
	\let \newlabelfcn=#4%
	\ifx \newlabelfcn\relax \else
	    \expandafter \newlabelfcn \csname#1\endcsname {#1}%
	\fi
    \fi
    \xdef \label  {\csname#1\endcsname}%
    \gdef \suffix {#2}%
    \ifx \suffix\empty \else
	\xdef \suffix {\expandafter\trimspace \suffix\end}%
	\xdef \label  {\label\suffix}%
    \fi
    }%
%
%
%
\newcount\appendixnumber	\appendixnumber=0
\newtoks\appendixstyle		\appendixstyle={\Alphabetic}
\newif\ifappendixlabel		\appendixlabelfalse
\def\APPEND#1{\par\penalty-300\vskip\chapterskip\spacecheck\chapterminspace
        \global\chapternumber=\number\appendixnumber
	\global\advance\appendixnumber by 1
	\chapterstyle\expandafter=\expandafter{\the\appendixstyle} \chapterreset
	\titlestyle{Appendix\ifappendixlabel~\chapterlabel\fi.~ {#1}}
	\nobreak\vskip\headskip\penalty 30000}
%

%
%
%

%
%
\newif\ifdraft\draftfalse
\newcount\yearltd\yearltd=\year\advance\yearltd by -1900
\def\draft{\drafttrue
	\def\draftdate{preliminary draft:
		\number\month/\number\day/\number\yearltd\ \ \hourmin}%
	\paperheadline={\hfil\draftdate} \headline=\paperheadline
	{\count255=\time\divide\count255 by 60 \xdef\hourmin{\number\count255}
        	\multiply\count255 by-60\advance\count255 by\time
		\xdef\hourmin{\hourmin:\ifnum\count255<10 0\fi\the\count255} }
	\message{draft mode}  }
%

\catcode`\@=12 
%


\refdef{adlerinducedgrav}{S. L. Adler,
{\sl Phys. Rev.} {\bf D14} (1976) 379; {\sl Phys. Rev. Lett.}
{\bf 44} (1980) 1567; {\sl Phys. Lett.}
{\bf 95B} (1980) 241; {\sl Rev. Mod. Phys.}
{\bf 54} (1982) 729.}
\refdef{alessandrinialo}{V. Alessandrini, D. Amati,
M. LeBellac, D. Olive,
{\sl Phys. Rep.} {\bf 1} (1971) 269.}
\refdef{alvarez}{O. Alvarez, {\sl Nucl. Phys.} {\bf B216} (1983) 125.}
\refdef{agaumegmv}{L. Alvarez-Gaume, C. Gomez, G. Moore, and C. Vafa,
{\sl Nucl. Phys.} {\bf B303} (1988) 455;
L. Alvarez-Gaume, C. Gomez, P. Nelson, G. Sierra, and C. Vafa,
{\sl Nucl. Phys.} {\bf B311} (1988) 333.}
\refdef{arefevam}{I. Ya. Arefeva and P. B. Medvedev,
{\sl Phys. Lett.} {\bf 202B} (1988) 510;
{\sl Phys. Lett.} {\bf 212B} (1988) 299.}
\refdef{amatibg}{D. Amati, C. Bouchiat, and J. L. Gervais,
{\sl Nuovo Cim. Lett.} {\bf 2} (1969) 399.}
\refdef{arnowittdm}{R. Arnowitt, S. Deser, and C. W. Misner,
{\sl Phys. Rev.} {\bf 117} (1960) 1595.}
\refdef{atickms}{J. J. Atick, G. Moore, and A. Sen,
SLAC preprint, SLAC-PUB 4463; {\sl Nucl. Phys.} {\bf B306} (1988) 279;
{\sl Nucl. Phys.} {\bf B308} (1988) 1.}
\refdef{awada}{M. A. Awada,
{\sl Phys. Lett.} {\bf 172B} (1986) 32.}
\refdef{ballestrero}{A. Ballestrero and E. Maina,
{\sl Phys. Lett.} {\bf 180B} (1986) 53;
{\sl Phys. Lett.} {\bf 182B} (1986) 317.}
\refdef{banksdss}{T. Banks, M. R. Douglas, N. Seiberg, and S. H. Shenker
Microscopic and Macroscopic Loops in Non-Perturbative
Two Dimensional Gravity,  Rutgers preprint RU-89-50.
}
\refdef{banksfmpp}{T. Banks, D. Friedan, E. Martinec,
M. Peskin, and C. Preitschopf,
{\sl Nucl. Phys.} {\bf B274} (1986) 71.}
\refdef{banksm}{T. Banks and E. Martinec,
private communication.}
\refdef{banksmrengrp}{T. Banks and E. Martinec,
{\sl Nucl. Phys.} {\bf B294} (1987) 733.}
\refdef{bankspesk}{T. Banks and M. Peskin,
{\sl Nucl. Phys.} {\bf B264} (1986) 513.}
\refdef{banksworm}{T. Banks,
{\sl Nucl. Phys.} {\bf B309} (1988) 493.}
\refdef{bardakcih}{K. Bardakci and M. B. Halpern,
{\sl Phys. Rev.} {\bf D3} (1971) 2493.}
\refdef{bardakcis}{K. Bardakci and S. Samuel,
{\sl Phys. Rev.} {\bf D16} (1977) 2500.}
\refdef{bardakcihs}{K. Bardakci, M. B. Halpern,
and J. A. Shapiro,
{\sl Phys. Rev.} {\bf 185} (1969) 1910.}
\refdef{batalink}{I. A. Batalin and R. E. Kallosh,
{\sl Nucl. Phys.} {\bf B222} (1983) 139.}
\refdef{batalinvopbrst}{I. A. Batalin and G. A. Vilkovisky,
{\sl Phys. Lett.} {\bf 69B} (1977) 309.}
\refdef{batalinvlagrangebrst}{I. A. Batalin and G. A. Vilkovisky,
{\sl Phys. Lett.} {\bf 102B} (1981) 27;
{\sl Phys. Rev.} {\bf D28} (1983) 2567.}
\refdef{bateman}{Bateman Manuscript Project, A. Erd{\'e}lyi, Editor,
 {\it Higher Transcendental
Functions}, vol. 2 (McGraw Hill, New York, 1953).}
\refdef{belavinpz}{A. A. Belavin,  A. M. Polyakov,  and
A. B. Zamolodchikov,
{\sl Nucl. Phys.} {\bf B241} (1984) 333}
\refdef{becchirs}{C. Becchi, A. Rouet, and R. Stora,
{\sl Phys. Lett.} {\bf 52B} (1974) 344;
{\sl Ann. Phys.} {\bf 98} (1976) 287.}
\refdef{bluhms}{R. Bluhm and S. Samuel,
``On the Modified Superstring Field Theory'',
preprint CCNY-HET-89/10, IUHET 170.}
\refdef{boch}{M. Bochicchio,
{\sl Phys. Lett.} {\bf 193B} (1987) 31.}
\refdef{boulwared}{D. G. Boulware and S. Deser,
{\sl Phys. Lett.} {\bf 40B} (1972) 227.}
\refdef{boulatovkkm}{D. V. Boulatov, V. A. Kazakov, I. K. Kostov,
and A. A> Migdal,
{\sl Nucl. Phys.} {\bf B275} (1986) 641.}
\refdef{braatencgt}{E. Braaten, T. Curtright, G. Ghandour, and C. Thorn,
{\sl Phys. Rev. Lett.} {\bf 51} (1983) 19;
{\sl Ann. Phys. (NY)} {\bf 153} (1984) 147.}
\refdef{braatenct}{E. Braaten, T. Curtright, and C. Thorn,
{\sl Phys. Lett.} {\bf 118B} (1982) 115;
{\sl Ann. Phys. (NY)} {\bf 147} (1983) 365.}
\refdef{brezinizp}{E. Brezin, C. Itzykson, G. Parisi, and J.-B. Zuber,
{\sl Comm. Math. Phys.} {\bf 59} (1978) 35.}
\refdef{brezink}{E. Brezin and V. A. Kazakov,
{\sl Phys. Lett.} {\bf 236B} (1990) 144.}
\refdef{brezinkz}{E. Brezin, V. A. Kazakov, and Al. B. Zamolodchikov,
{\sl Nucl. Phys.} {\bf B338} (1990) 673.
}
\refdef{brezindks}{E. Brezin, M. R. Douglas, V. Kazakov, and S. H. Shenker,
The Ising Model Coupled to 2D Gravity:A Nonperturbative Analysis,
Rutgers preprint, RU-89-47.}
\refdef{brinknielsen}{L. Brink and Nielsen, H.,
{\sl Phys. Lett.} {\bf 43B} (1973) 319;
{\sl Nucl. Phys.} {\bf B89} (1975) 18.}
\refdef{browerf}{R.C. Brower and K.A. Friedman,
{\sl Phys. Rev.} {\bf D7}  (1973) 535. }
\refdef{browerg}{R. C. Brower and P. Goddard,
{\sl Lett. al Nuovo Cimento} {\bf 1} (1971) 1075.}
\refdef{browert}{R.C. Brower and C. B. Thorn,
{\sl Nucl. Phys.} {\bf B31} (1971)163. }
\refdef{brower}{R. C. Brower,
{\sl Phys. Rev.} {\bf D6} (1972) 1655.}
\refdef{callanlny}{C. G. Callan, C. Lovelace,  C. R. Nappi, and
S. A. Yost,
{\sl Nucl. Phys.} {\bf 288} (1987) 525; {\bf B293} (1987) 83.}
\refdef{candelashsw}{P. Candelas, G. Horowitz, A. Strominger, and E. Witten,
{\sl Nucl. Phys.} {\bf B258} (1985) 46.}
\refdef{capelliiz}{Cappelli, A., Itzykson, C., and
Zuber, J.-B.,
{\sl Comm. Math. Phys.} {\bf 113} (1987) 1.}
\refdef{cardy}{Cardy, J.,
{\sl Nucl. Phys.} {\bf B270} (1986) 186.}
\refdef{chodost}{A. Chodos and C. B. Thorn,
{\sl Nucl. Phys.} {\bf B72} (1974) 509.}
\refdef{choihworm}{K. Choi and R. Holman,
{\sl Phys. Rev. Lett.} {\bf 62} (1989) 2575.}
\refdef{clavellis}{L. Clavelli and J. A. Shapiro,
{\sl Nucl. Phys.} {\bf B57} (1973) 490.}
\refdef{coleman}{S. Coleman,
{\sl Phys. Rev.} {\bf D11} (1975) 2088.}
\refdef{colemanworm1}{S. Coleman,
{\sl Nucl. Phys.} {\bf B307} (1988) 867.}
\refdef{colemanworm2}{S. Coleman,
{\sl Nucl. Phys.} {\bf B310} (1988) 643.}
\refdef{collins}{J. C. Collins,
{\it Renormalization}, Cambridge Univ. Press, Cambridge, 1984.}
\refdef{coon}{D. D. Coon,
{\sl Phys. Lett.} {\bf 29B} (1969) 669.}
\refdef{coonb}{M. Baker and D. D. Coon,
{\sl Phys. Rev.} {\bf D13} (1976) 707.}
\refdef{corrigangnogh}{E. F. Corrigan and P. Goddard,
{\sl Nucl. Phys.} {\bf B68} (1974) 189.}
\refdef{corriganofermions}{E. F. Corrigan and D. Olive,
{\sl Nuovo Cim.} {\bf 11A} (1972) 749.}
\refdef{cremmerg}{E. Cremmer and J. L. Gervais,
{\sl Nucl. Phys.} {\bf B76} (1974) 209;
{\bf B90} (1975) 410.}
\refdef{cremmers}{E. Cremmer and J. Scherk,
{\sl Nucl. Phys.} {\bf B50} (1972) 222.}
\refdef{cremmerscurrents}{E. Cremmer and J. Scherk,
{\sl Nucl. Phys.} {\bf B48} (1972) 29.}
\refdef{cremmerst}{E. Cremmer, A. Schwimmer, and C. B. Thorn,
{\sl Phys. Lett.} {\bf 179B} (1986) 57; see also
C. B. Thorn,
{\it Proceedings of the
XXIII International Conference on High Energy Physics, 16-23 July 1986,
Berkeley, California}, ed. Stuart Loken, World Scientific (1987), p. 374.}
\refdef{crewtherdvw}{R. J. Crewther, P. diVecchia, G. Veneziano,
and E. Witten,
{\sl Phys. Lett.} {\bf 88B} (1979) 123.}
\refdef{curtrighttliou}{T. L. Curtright and C. B. Thorn,
{\sl Phys. Rev. Lett.} {\bf 48} (1982) 1309.}
\refdef{curtrightt}{T. L. Curtright and C. B. Thorn,
{\sl Nucl. Phys.} {\bf B274} (1986) 520.}
\refdef{curtrightgt}{T. L. Curtright, C. B. Thorn, and J. Goldstone,
{\sl Phys. Lett.} {\bf 175B} (1986) 47.}
\refdef{dashen}{R. Dashen, {\sl Phys. Rev.} {\bf D3} (1971) 1879.}
\refdef{david}{F. David,
{\sl Mod. Phys. Lett.} {\bf A2} (1988) 1651.}
\refdef{dailp}{J. Dai, R. G. Leigh, J. Polchinski,
{\sl Mod. Phys. Lett.} {\bf A4} (1989) 2073.}
\refdef{delgiudicedf}{E. Del Giudice, P. di Vecchia, and S. Fubini,
{\sl Ann. Phys.} {\bf 70} (1972) 378.}
\refdef{deser}{S. Deser,
{\sl Ann. Phys.} {\bf 76} (1973) 189.}
\refdef{dewitt}{B. S. DeWitt,
{\sl Phys. Rev.} {\bf 162} (1967) 1195.}
\refdef{dhokerg}{E. D'Hoker and S. B. Giddings,to be published,
{\sl Nucl. Phys.}}
\refdef{dhokerp}{E. D'Hoker and D. H. Phong,
{\sl Phys. Rev. Lett.} {\bf 56} (1986) 912;
{\sl Nucl. Phys.} {\bf B269} (1986)205;
{\sl Rev. Mod. Phys.} {\bf 60} (1988) 917.}
\refdef{distlerk}{J. Distler and H. Kawai,
{\sl Nucl. Phys.} {\bf B321} (1989) 509.}
\refdef{dixonhvw}{L. Dixon, J. Harvey, C. Vafa, and E. Witten,
{\sl Nucl. Phys.} {\bf B261} (1985) 678.}
\refdef{dotsenkof}{Vl.S. Dotsenko and V.A. Fateev,
{\sl Nucl. Phys.} {\bf B240} [FS12] (1984) 312.}
\refdef{douglas}{M. R. Douglas,
Strings in Less than One Dimension and the Generalized
KDV Hierarchies, rutgers preprint, RU-89-51.}
\refdef{douglass}{M. R. Douglas and S. H. Shenker,
{\sl Nucl. Phys.} {\bf B335} (1990) 635.}
\refdef{duffps}{M. J. Duff, C. N. Pope, and K. S. Stelle,
{\sl Phys. Lett.} {\bf B223} (1989) 386.}
\refdef{fairlie}{D. Fairlie, unpublished.}
\refdef{feigin}{B. L. Feigin, and D. B. Fuchs,
Funkts. Anal. Prilozhen. {\bf 16} (1982) 47
[Funct. Anal. Appl. {\bf 16} (1982) 114];
Funkts. Anal. Prilozhen. {\bf 17} (1983) 91.}
\refdef{fischlers}{Fischler, W., L. Susskind,
{\sl Phys. Lett.}{\bf 171B} (1986) 383;
{\bf 173B} (1986) 262.}
\refdef{fradkinfbrst}{E. S. Fradkin and T. E. Fradkina,
{\sl Phys. Lett.}{\bf 72B} (1978) 343.}
\refdef{fradkinvbrst}{E. S. Fradkin and G. A. Vilkovisky,
{\sl Phys. Lett.} {\bf 55B} (1975) 224.}
\refdef{freedmangst}{D. Z. Freedman,  S. B. Giddings,
Shapiro, J. A., and C. B. Thorn,
{\sl Nucl. Phys.} {\bf B298} (1988) 253.}
\refdef{freemano}{M.D. Freeman, D. Olive,
{\sl Phys. Lett.} {\bf B175} (1986)151.}
\refdef{frenkelgz}{I.B. Frenkel, H. Garland, and
G.J. Zuckerman, Yale Univ., Dept. of Math., preprint (1986).}
\refdef{frenkelk}{I.B. Frenkel and V.G. Kac,
{\sl Inv. Math.} {\bf 62} (1980) 23.}
\refdef{freundw}{P. G. O. Freund and E. Witten,
{\sl Phys. Lett.} {\bf 199B} (1987) 191.}
\refdef{friedansft}{D. Friedan,
{\sl Nucl. Phys.} {\bf B271} (1986) 540.}
\refdef{friedanqs}{Friedan, D., Qiu, Z., and Shenker, S. H.,
{\sl Phys. Rev. Lett.}{\bf 52} (1984) 1575.}
\refdef{friedanms}{D. Friedan, E. Martinec, and S. Shenker,
{\sl Phys. Lett.} {\bf 160B} (1985) 55;
{\sl Nucl. Phys.} {\bf B271} (1986) 93.}
\refdef{friedans}{D. Friedan and S. Shenker,
{\sl Phys. Lett.} {\bf 175B} (1986) 287.}
\refdef{fubiniv}{S. Fubini and G. Veneziano,
Ann. Phys. {\bf 63} (1971) 12.}
\refdef{fubinigv}{S. Fubini, D. Gordon, and G. Veneziano,
{\sl Phys. Lett.} {\bf29B}(1969) 679.}
\refdef{fujikawa}{K. Fujikawa,
{\sl Phys. Rev.} {\bf D25} (1982) 2584.}
\refdef{gellmannglmz}{M. Gell-Mann, M. L. Goldberger,
F. E. Low, E. Marx, and F. Zachariasen,
{\sl Phys. Rev.} {\bf 133} (1964) B145.}
\refdef{gellmannor}{M. Gell-Mann, R. J. Oakes, and B. Renner,
{\sl Phys. Rev.} {\bf 175} (1968) 2195.}
\refdef{gervaisnliou}{
J. L. Gervais and A. Neveu,
          {\sl Nucl. Phys.} {\bf B199} (1982) 59, {\bf B209} (1982) 125,
          {\sl Phys. Lett.} {\bf 123B} (1983) 86, {\sl Nucl. Phys.}
       {\bf B224} (1983) 329;   
    Lecture by Gervais in {\it Perspectives in String Theory},
    ed. P. diVecchia and J. L. Petersen, World Scientific, 1988.}
\refdef{gervais}{J.-L. Gervais and A. Neveu,
{\sl Phys. Lett.} {\bf 151B} (1985) 271;
J.-L. Gervais,
See also lectures by Gervais in the
last of \Refer{gervaisnliou}.}
\refdef{giddingsfour}{S. B. Giddings,
{\sl Nucl. Phys.} {\bf B278}(1986) 242.}
\refdef{giddingssd}{S. B. Giddings, in proceedings of the conference
``Mathematical Aspects of   String Theory'',
ed. S.-T. Yau, World Scientific, in press.}
\refdef{giddingsmw}{S. B. Giddings, E. Martinec, and E. Witten,
{\sl Phys. Lett.} {\bf 176B} (1986) 362.}
\refdef{giddingsm}{S. B. Giddings and E. Martinec,
{\sl Nucl. Phys.} {\bf B278} (1986) 91.}
\refdef{giddingssworm}{S. B. Giddings and A. Strominger,
{\sl Nucl. Phys.} {\bf B307} (1988) 854.}
\refdef{gilesmt}{R. Giles, L. McLerran, C. B. Thorn,
{\sl Phys. Rev.} {\bf D17} (1978) 2058.}
\refdef{gilest}{R. Giles and C. B. Thorn,
{\sl Phys. Rev.} {\bf D16} (1977) 366.}
\refdef{gliozziso}{F. Gliozzi, J. Scherk, and D. Olive,
{\sl Phys. Lett.} {\bf 65B} (1976) 282;
{\sl Nucl. Phys.} {\bf B122} (1977) 253.}
\refdef{goddardo}{P. Goddard and D. Olive, in
{\sl Vertex Operators in Mathematics and Physics},
MSRI Publication \# 3 (Springer, Heidelberg, 1984) p. 51.}
\refdef{goddardrt}{P. Goddard, C. Rebbi, and C. B. Thorn,
{\sl Il Nuovo Cimento}, {\bf 12A} (1972) 425.}
\refdef{goddardgrt}{P. Goddard, J. Goldstone,
C. Rebbi, and C. B. Thorn, {\sl Nucl. Phys.} {\bf B56} (1973) 109.}
\refdef{goddardt}{P. Goddard and C. B. Thorn,
{\sl Phys. Lett.} {\bf 40B} (1972) 235.}
\refdef{gorshkovglf}{G. Gorshkov, V. N. Gribov,
L. N. Lipatov, and G. V. Frolov,
({\sl Yad. Fiz.} {\bf 6} (1967) 129)
{\sl Sov. Jour. of Nucl. Phys.} {\bf 6} (1968) 95.}
\refdef{gotostring}{ T. Goto,
Prog. Theor. Phys. {\bf 46} (1971) 1560.}
\refdef{greenmosc}{M. B. Green,
``Point-like Structure in String Theory,''
Invited talk at the First International A. D. Sakharov
Conference on Physics, 27--31 May, 1991,
Lebedev Institute, Moscow.}
\refdef{greensanomaly}{M. B. Green and J. H. Schwarz,
{\sl Phys. Lett.} {\bf 149B} (1984) 117;
{\sl Nucl. Phys.} {\bf B255} (1985) 93.}
\refdef{greensbook}{M. B. Green, J. H. Schwarz, and E. Witten,
{\it Superstring Theory}, Volumes 1 and 2,
Cambridge University Press (1987).}
\refdef{greensspinor}{M. B. Green and J. H. Schwarz,
{\sl Phys. Lett.} {\bf 136B} (1984) 367;
{\sl Nucl. Phys.} {\bf B243} (1984) 285.}
\refdef{greenssft}{M. B. Green and J. H. Schwarz,
{\sl Nucl. Phys.} {\bf B243} (1984) 475.}
\refdef{greenseiberg}{M. B. Green and N. Seiberg,
{\sl Nucl. Phys.} {\bf B299} (1988) 108.}
\refdef{greensiteklink}{J. Greensite and F. R. Klinkhamer,
{\sl Nucl. Phys.} {\bf B281} (1987) 269;
{\sl Nucl. Phys.} {\bf B291} (1987) 557;
{\sl Nucl. Phys.} {\bf B304} (1988) 108.}
\refdef{grosshmr}{D. J. Gross, J. A. Harvey, E. Martinec, and R. Rohm,
{\sl Phys. Rev. Lett.} {\bf 54} (1985) 502;
{\sl Nucl. Phys.} {\bf B256} (1985) 253;
{\sl Nucl. Phys.} {\bf B267} (1986) 75.}
\refdef{grossj}{D. Gross, and A. Jevicki,
{\sl Nucl. Phys.} {\bf B283} (1987) 1;
{\sl Nucl. Phys.} {\bf B287} (1987) 225.}
\refdef{grossjsup}{D. Gross, and A. Jevicki,
{\sl Nucl. Phys.} {\bf B293} (1987) 29. }
\refdef{grossm}{D. J. Gross and A. A. Migdal,
{\sl Phys. Rev. Lett.} {\bf 64} (1990) 127;
{\sl Phys. Rev. Lett.} {\bf 64} (1990) 717;
{\sl Nucl. Phys.} {\bf B340} (1990) 333.
{\sl Phys. Rev. Lett.} {\bf 64} (1990) 717.
}
\refdef{grossmilc}{D. J. Gross and N. Miljkovic,
{\sl Phys. Lett.} {\bf 238B} (1990) 217.}
\refdef{grossnss}{D. Gross, A. Neveu, J. Scherk, and J. H. Schwarz,
{\sl Phys. Rev.} {\bf D2} (1970) 697.}
\refdef{grossp}{D. Gross and V. Periwal,
{\sl Phys. Rev. Lett.} {\bf 60} (1988) 2105.}
\refdef{grosspy}{D. Gross, R. D. Pisarski, and L. G. Yaffe,
{\sl Rev. Mod. Phys.} {\bf 53} (1981) 43.}
\refdef{halpern}{M. Halpern,
{\sl Phys. Rev.} {\bf D12} (1975) 1684.}
\refdef{hamermesh}{M. Hamermesh,
{\it Group Theory}, Addison-Wesley, 1962.}
\refdef{hataikkoopen}{H. Hata, K. Itoh,
T. Kugo, H. Kunitomo, and K. Ogawa,
{\sl Phys. Lett.} {\bf 172B} (1986) 186;
{\sl Phys. Lett.} {\bf 172B} (1986) 195;
{\sl Phys. Rev.} {\bf D34} (1986) 2360;
Phys, Rev. {\bf D35} (1986) 1356.}
\refdef{hataikkopregeom}{H. Hata, K. Itoh,
T. Kugo, H. Kunitomo, and K. Ogawa,
{\sl Phys. Lett.} {\bf 175B} (1986) 138.}
\refdef{hataikkoclosed}{H. Hata, K. Itoh,
T. Kugo, H. Kunitomo, and K. Ogawa,
{\sl Phys. Rev.} {\bf D35} (1986) 1318;
Prog. Theor. Phys. {\bf B77} (1987) 443.}
\refdef{hawkingworm}{S. W. Hawking,
{\sl Phys. Lett.} {\bf 195B} (1987) 337.}
\refdef{horowitzmm}{G. T. Horowitz, S. P. Martin, and R. C. Meyers,
``Remarks on Superconformal Bosonization,''
Santa Barbara preprint (1988), NSF-ITP-88-112.}
\refdef{horowitzlrs}{G. T. Horowitz, J. Lykken, R. Rohm, A.Strominger,
{\sl Phys. Rev. Lett.} {\bf 57} (1986) 283.}
\refdef{horowitzs}{G. T. Horowitz and A. Strominger,
{\sl Phys. Lett.} {\bf 185B} (1987) 45.}
\refdef{hsuesakitav}{C. S. Hsue, B. Sakita, and M. A. Virasoro,
{\sl Phys. Rev.} {\bf D2} (1970) 2857.}
\refdef{hwang}{S. Hwang,
{\sl Phys. Rev.} {\bf D28}  (1983) 2614.}
\refdef{ishamss}{C. J. Isham, A. Salam, and J. Strathdee,
{\sl Lett. al Nuov. Cim.} {\bf 5} (1972) 969.}
\refdef{jacobi}{C. G. J. Jacobi,
{\it Fundamenta Nova Theoriae Functionum Ellipticarum},
Konigsberg, 1829, p. 145.}
\refdef{johnson}{K. Johnson,
{\it Lectures on Particles and Field Theory},
S. Deser and K. W. Ford, Editors
Prentice-Hall (1964).}
\refdef{mkac}{M. Kac,
Amer. Math. Monthly, {\bf 73}, part II, (1966) 1.}
\refdef{kacdet}{V. G. Kac,
in {\it Group Theoretical Methods in Physics}, edited by
W. Beiglb\"ock, A. B\"ohm, and E. Takasugi (Springer-Verlag,
New York, 1979, p. 441. For the proof of the
formula, see Feigin and Fuchs\[feigin].}
\refdef{kakuthorn}{M. Kaku and C. B. Thorn,
{\sl Phys. Rev.} {\bf D1} (1970) 2860.}
\refdef{kakuyu}{M. Kaku and L. Yu,
{\sl Phys. Lett.} {\bf 33B} (1970) 166;
{\sl Phys. Rev.} {\bf D3} 2992, 3007.}
\refdef{kakukikkawa}{M. Kaku and K. Kikkawa,
{\sl Phys. Rev.} {\bf D10} (1974) 1823.}
\refdef{kalloshghost}{R. E. Kallosh,
{\sl Nucl. Phys.} {\bf B141} (1978) 141.}
\refdef{kalloshgrsch}{R. E. Kallosh,
{\sl Phys. Lett.} {\bf 195B} (1987) 369.}
\refdef{kalloshm}{R. E. Kallosh and A. Morozov,
{\sl Int. Jour. of Mod. Phys.} {\bf A3} (1988) 1943.}
\refdef{kalloshr}{R. E. Kallosh and M. A. Rahmanov,
``Covariant Quantization of the Green
Schwarz Superstring,'' Lebedev preprint (1988), 88-0448.}
\refdef{kalloshtvp}{R. E. Kallosh, W. Troost, and A. Van Proeyen,
``Quantization of Superparticle and Superstring
with Siegel's Modification,''
Leuven preprint(1988), KUL-TF-88/9.}
\refdef{kazakov}{V. A. Kazakov,
{\sl Mod. Phys. Lett.} {\bf A4} (1989) 2125.}
\refdef{kazakovm}{V. A. Kazakov and A. A. Migdal,
{\sl Nucl. Phys.} {\bf B311} (1988/89) 171.}
\refdef{katoogawa}{M. Kato and K. Ogawa,
{\sl Nucl. Phys.} {\bf B212} (1983) 443.}
\refdef{kikkawasv}{K. Kikkawa, B. Sakita, and M. A. Virasoro,
{\sl Phys. Rev.} {\bf 184} (1969) 1701.}
\refdef{kirschnerlnonvac}{R. Kirschner and L. N. Lipatov,
({\sl Zh. Eksp. Teor. Fiz.} {\bf 83} (1982) 488)
{\sl Sov. Phys. JETP} {\bf 56} (1982) 266;
see also J. Kwiecinski,
{\sl Phys. Rev.} {\bf D26} (1982) 3293.}
\refdef{kirschner}{R. Kirschner,
{\sl Z. Phys. C - Particles and Fields} {\bf 31} (1986) 135.}
\refdef{kirschnerlbarereggeons}{R. Kirschner and L. N. Lipatov,
{\sl Z. Phys. C - Particles and Fields} {\bf 45} (1990) 477.}
\refdef{knizhnik}{V. G. Knizhnik,
{\sl Phys. Lett.} {\bf 160B} (1985) 403.}
\refdef{knizhnikpz}{V. G. Knizhnik, A. M. Polyakov, and A. B.
Zamolodchikov, {\sl Mod. Phys. Lett.} {\bf A2} (1988) 819.}
\refdef{koban}{Z. Koba and H. B. Nielsen,
{\sl Nucl. Phys.} {\bf B12} (1969) 517.}
\refdef{kugoojima}{T. Kugo and I. Ojima, {\sl Phys. Lett.} {\bf 73B}
(1978) 459; Prog. Theor. Phys. Supplement {\bf 66} 1979 1.}
%
\refdef{kugolectures}{T. Kugo, ``Covariantized Light-Cone
String Field Theory'', Kyoto preprint, KUNS 917 HE (TH) 88/02.}
\refdef{kugoterao}{ T. Kugo and H. Terao,
{\sl Phys. Lett.} {\bf 208B} (1988) 416.}
%
\refdef{kugotu}{T. Kugo, H. Terao, and Uehara,
{\sl Supp. Prog. Theor. Phys.} {\bf 85} (1985) 122.}
\refdef{kunitomos}{H. Kunitomo and K. Suehiro,
{\sl Nucl. Phys.} {\bf B289} (1987) 157;
``Operator Expression of Witten's Superstring Vertex'',
Kyoto preprint KUNS-857 (1987).}
\refdef{kwiecinski}{J. Kwiecinski,
{\sl Phys. Rev.} {\bf D26} (1982) 3293.}
\refdef{lancaster}{D. Lancaster,
{\sl Phys. Lett.} {\bf 203B} (1988) 224.}
%
\refdef{leclairpp}{A. LeClair, M. E. Peskin, and C. R. Preitschopf,
{\sl Nucl. Phys.} {\bf B317} (1989) 411;
{\sl Nucl. Phys.} {\bf B317} (1989) 464;
C. R. Preitschopf,
The Gluing Theorem in the Operator Formulation of
String Field Theory, in {\it Superstrings}, edited
by P. G. O. Freund and K. T. Mahanthappa, p. 39, Plenum (1988);
A. LeClair,
{\sl Nucl. Phys.} {\bf B297}, (1988) 603;
{\sl Nucl. Phys.} {\bf B303}, (1988) 189.}
\refdef{lechtenfeld}{O. Lechtenfeld and S. Samuel,
{\sl Nucl. Phys.} {\bf B310} (1988) 254.}
\refdef{lechtenfeldsmodification}{O. Lechtenfeld and S. Samuel,
`` Gauge-invariant Modification of Witten's Open Superstring,''
City College of New York preprint CCNY-HEP-88/8.}
\refdef{lipatovpomeron}{L. N. Lipatov,
({\sl Zh. Eksp. Teor. Fiz.} {\bf 90} (1986) 1536)
{\sl Sov. Phys. JETP} {\bf 63} (1986) 904.}
\refdef{lovelaceloops}{C. Lovelace,
{\sl Phys. Lett.} {\bf 32B} (1970) 703.}
\refdef{lovelacecrit}{C. Lovelace,
{\sl Phys. Lett.} {\bf 34B} (1971) 500.}
\refdef{lovelacefreeregge}{C. Lovelace, 
{\sl Nucl. Phys.} {\bf B95} (1975) 12.}
\refdef{lovelace}{C. Lovelace,
{\sl Nucl. Phys.} {\bf B273} (1986) 413.}
\refdef{lykkenr}{J. Lykken and S. Raby,
{\sl Nucl. Phys.} {\bf B278} (1986) 256.}
\refdef{mckeans}{H. P. McKean and I. M. Singer,
J. Diff. Geom. {\bf 1} (1967) 43.}
\refdef{mandelstam}{S.Mandelstam,
{\sl Nucl. Phys.} {\bf B64} (1973) 205;
{\sl Phys. Lett.} {\bf 46B} (1973) 447;
{\sl Nucl. Phys.} {\bf B69} (1974) 77;
see also his lectures in
{\it Unified String Theories}, ed. M. Green and D. Gross
(World Scientific) 1986.}
\refdef{mandelstambose}{S. Mandelstam,
{\sl Phys. Rev.} {\bf D11} (1975) 3026.}
\refdef{mandelstambrandeis}{S. Mandelstam, in
{\it Lectures on Elementary Particles and Quantum Field Theory},
edited by S. Deser, M. Grisaru, and H. Pendleton,
(MIT Press, Cambridge, Mass., 1970).}
\refdef{mandelstamlorentz}{S. Mandelstam,
{\sl Nucl. Phys.} {\bf B83} (1974) 413.}
\refdef{marnelius}{R. Marnelius,
{\sl Nucl. Phys.} {\bf B211} (1983) 14.}
\refdef{marneliusspinliou}{R. Marnelius,
{\sl Nucl. Phys.} {\bf B221} (1983) 409.}
\refdef{marneliusliou}{
R. Marnelius, {\sl Nucl. Phys.} {\bf B211}
          (1983) 14, {\sl Phys. Lett.} {\bf 123B} (1983) 237;
          Lecture in {\it Perspectives in String Theory},
    ed. P. diVecchia and J. L. Petersen, World Scientific, 1988.}
\refdef{marshallr}{C. Marshall and P. Ramond,
{\sl Nucl. Phys.} {\bf B85} (1975) 375.}
\refdef{martinec}{E. Martinec, private communication.}
\refdef{mcclainr}{B. McClain and B. Roth,
Comm. Math. Phys. {\bf 111} (1987) 539; see also
{\sl Phys. Rev.} {\bf D36} (1987) 1184.}
\refdef{mooren}{G. Moore and P. Nelson,
{\sl Nucl. Phys.} {\bf B266} (1986) 58.}
\refdef{nahm}{W. Nahm,
{\sl Nucl. Phys.} {\bf B81} (1974) 164;
{\bf B114} (1976) 174;
{\bf B124} (1977) 121;
{\bf B120} (1977) 125.}
\refdef{nakanishio}{N. Nakanishi and I. Ojima,
{\sl Phys. Rev. Lett.} {\bf 43} (1979) 91.}
\refdef{nambustring}{Y. Nambu,
Lectures at the Copenhagen Symposium (1970).}
\refdef{neveuk}{A. Neveu and Y. Kazama,
{\sl Nucl. Phys.} {\bf B276} (1986) 366.}
\refdef{neveunw}{A. Neveu, H. Nicolai, and P. C. West,
{\sl Phys. Lett.} {\bf 167B} (1986) 307}
\refdef{neveuscurrents}{A. Neveu and J. Scherk,
{\sl Nucl. Phys.} {\bf B41} (1972) 365.}
\refdef{neveus}{A. Neveu and J. H. Schwarz,
{\sl Nucl. Phys.} {\bf B31} (1971) 86.}
\refdef{neveusferm}{A. Neveu and J. H. Schwarz,
{\sl Phys. Rev.} {\bf D4} (1971) 1109.}
\refdef{neveust}{A. Neveu, J. H. Schwarz, and C. B. Thorn,
{\sl Phys. Lett.} {\bf 35B} (1971) 529.}
\refdef{neveuw}{A. Neveu and P. West,
{\sl Phys. Lett.} {\bf 168B} (1986) 192;
{\sl Phys. Lett.} {\bf 179B} (1986) 235.}
\refdef{nielsennworm}{H. B. Nielsen and M. Ninomiya,
U. of Tokyo preprint INS-Rep-727 (Dec. 1988).}
\refdef{nielsenghost}{N. K. Nielsen,
{\sl Nucl. Phys.} {\bf B140} (1978) 599.}
\refdef{nielsenfishnet}{H. B. Nielsen and P. Olesen,
{\sl Phys. Lett.} {\bf 32B} (1970) 203;
B. Sakita and M. A. Virasoro,
{\sl Phys. Rev. Lett.} {\bf 24} (1970) 1146.}
\refdef{nissinovp}{E. R. Nissimov and S. J. Pacheva,
{\sl Phys. Lett.} {\bf 202B} (1988) 325.}
\refdef{nissinovps}{E. R. Nissimov. S. J. Pacheva, and
S. Solomon,
{\sl Nucl. Phys.} {\bf B297} (1988) 369;
see also {\it Perspectives in String Theory},
Ed. P. DiVecchia and J. L. Petersen,
World Scientific (1988), p. 182.}
\refdef{ogievetsky}{V. I. Ogievetsky,
{\sl Lett. al Nuovo Cimento} {\bf 8} (1973) 988.}
\refdef{parisis}{G. Parisi and N. Sourlas,
{\sl Phys. Rev. Lett.} {\bf 43} (1979) 744.}
\refdef{pauli}{W. Pauli and M. Fierz,
{\sl Proc. Roy. Soc.} {\bf A173} (1939) 211.}
\refdef{peskin}{M. E. Peskin, private communication.}
\refdef{peskinthorn}{M. E. Peskin,  and C. B. Thorn,
{\sl Nucl. Phys.} {\bf B269} (1986) 509.}
\refdef{petrovs}{V. A. Petrov and A. P. Samokhin,
{\sl Phys. Lett.} {\bf 237B} (1990) 500.}
\refdef{pfeffer}{D. Pfeffer, P. Ramond, and V. C. J. Rodgers,
{\sl Nucl. Phys.} {\bf B276} (1986) 131.}
\refdef{polchinskihotstring}{J. Polchinski,
Comm. Math. Phys. {\bf 104} (1986) 37.}
\refdef{polchinski2dgrav}{J. Polchinski,
{\sl Nucl. Phys.} {\bf B324} (1989) 123.}
\refdef{polyakov}{A. M. Polyakov,
{\sl Phys. Lett.} {\bf 103B} (1981) 207.}
\refdef{polyakovsup}{A. M. Polyakov,
{\sl Phys. Lett.} {\bf103B} (1981) 211. }
\refdef{polyakovlightcone}{A. M. Poluakov,
{\sl Mod. Phys. Lett.} {\bf A2} (1987) 893.}
\refdef{preitschopf}{C. R. Preitschopf,
``Superconformal Ghosts: Chiral Bosonization
at Genus 0 and 1,''
University of Maryland preprint, 88-206
(1988).}
\refdef{preitschopfty}{C.R. Preitschopf, C.B. Thorn and S.A. Yost,
{\sl Nucl. Phys.} {\bf B337} (1990) 363-433.}
\refdef{preitschopftytusca}{C.R. Preitschopf, C.B. Thorn and S.A. Yost,
``Superstring Field Theory,''
University of Florida preprint UFTP-HEP-90-3,
invited lecture
published in {\it Superstrings
and Particle Theory}, Proc. of
a Workshop at Tuscaloosa Alabama,
8-11 November, 1989, L. Clavelli and B. Harms, Eds.,
World Scientific (1990).}
\refdef{preitschopftsubsft}{C.R. Preitschopf, C.B. Thorn,
``Action Principle for Subcritical String Fields,''
University of Florida preprint UFIFT-HEP-90-17,
to be published in
{\sl Nucl. Phys.}{\bf B349} (1991) 132. }
\refdef{preitschopftsubback}{C.R. Preitschopf, C.B. Thorn,
{\sl Phys. Lett.}{\bf 250B} (1990) 79. }
\refdef{preskilltwworm}{J. Preskill, S. P. Trivedi, and M. Wise,
{\sl Phys. Lett.} {\bf 223B} (1989) 26.}
\refdef{qius}{Z. Qiu and A. Strominger,
{\sl Phys. Rev.} {\bf D36} (1987) 1794.}
\refdef{ramond}{P. Ramond,
{\sl Phys. Rev.} {\bf D3} (1971) 2415.}
\refdef{ramondqaction}{P. Ramond,
{\sl Prog. Theor. Suppl.} {\bf 86} (1986) 126.}
\refdef{romans}{L. J. Romans,
{\sl Phys. Lett.} {\bf 194B} (1987) 499.}
%
\refdef{sakharov}{A. D. Sakharov,
{\sl Dok. Akad. Nauk. SSSR} {\bf 177} (1967) 70
[{\sl Sov. Phys. Dokl.} {\bf 12} (1968) 1040].}
\refdef{samuel}{S. Samuel,
{\sl Phys. Lett.} {\bf 181B} (1986) 249;
{\sl Phys. Lett.} {\bf 181B} (1986) 255;
{\sl Nucl. Phys.} {\bf B296} (1988) 187.}
\refdef{sazdovic}{B. Sazdovic,
{\sl Phys. Lett.} {\bf 195B} (1987) 536.}
\refdef{schwarznogh}{J. Schwarz,
{\sl Nucl. Phys.} {\bf B46} (1972) 61.}
\refdef{segal}{G. Segal, {\sl Comm. Math. Phys.} {\bf 80}
(1981) 301.}
\refdef{sen}{S. Sen and R. Holman,
{\sl Phys. Rev. Lett.} {\bf 58} (1987) 1304.}
\refdef{shapirotwups}{J. A. Shapiro and C. B. Thorn,
 {\sl Phys. Lett.} {\bf 194B} (1987) 43.}
\refdef{shapirotcups}{J. A. Shapiro and C. B. Thorn,
{\sl Phys. Rev.} {\bf D36} (1987) 432.}
\refdef{shapiro}{Shapiro, J. A.,
{\sl Phys. Rev.} {\bf D5} (1972) 1945.}
\refdef{siegelbrst}{W. Siegel,
{\sl Phys. Lett.} {\bf 142B} (1984) 276;
{\sl Phys. Lett.} {\bf 149B} (1984) 157, 162;
{\bf 151B} (1985) 391, 396.}
\refdef{siegelzunfix}{W. Siegel and B. Zwiebach,
{\sl Nucl. Phys.} {\bf B263} (1986) 105.}
\refdef{siegelzwie}{W. Siegel and B. Zwiebach,
Nucl Phys. {\bf B282} (1987) 125; 
{\sl Phys. Lett.} {\bf 184B} (1987) 325;
{\sl Nucl. Phys.} {\bf 288} (1987) 332;
{\sl Nucl. Phys.} {\bf 299} (1988) 206.}
\refdef{sikiviet}{P. Sikivie and C. B. Thorn,
{\sl Phys. Lett.} {\bf 234B} (1990) 132-134.}
\refdef{spiegelglas}{M. Spiegelglas,
IAS preprint (1986). }
\refdef{strominger}{A. Strominger,
{\sl Phys. Rev. Lett.} {\bf 58} (1987)629;
 Nucl.Phys.{\bf B294} (1987) 93.}
\refdef{stronggravity}{V. I. Ogievetsky and I. V. Polubarinov,
{\sl Annals of Physics} {\bf 35} (1965) 167;
B. Zumino, {\it Lectures on Elementary Particles and Quantum
Field Theory}. 1970 Brandeis Summer Inst. in Theor. Phys.,
Ed. S. Deser, M. Grisaru, and H. Pendleton
(MIT Press, 1970);
C. J. Isham, A. Salam, and J. Strathdee,
{\sl Phys. Rev.} {\bf D3} (1971) 867.}
\refdef{stueckelberg}{E. C. G. Stueckelberg,
{\sl Helv. Phys. Acta} {\bf 30} (1957) 209.}
\refdef{suehiro}{K. Suehiro,
{\sl Prog. Theor. Phys.} {\bf 78} (1987) 1151;
{\sl Nucl. Phys.} {\bf B296} (1988) 333.}
%
\refdef{thooft}{G. `t Hooft,
in {\it Recent Developments in
Gauge Theories}, G. `t Hooft {\it et al}, Eds.
(Plenum, New York, 1980).}
\refdef{thooftlargen}{G. 't Hooft,
{\sl Nucl. Phys.} {\bf B72} (1974) 461.}
\refdef{thooftfishnet}{G. 't Hooft,
{\sl Nucl. Phys.} {\bf B} (1974) .}
\refdef{thornchiral}{C.B. Thorn,
{\sl Phys. Rev.} {\bf D23} (1981) 439.}
\refdef{thornnonlinear}{C.B. Thorn,
{\sl Phys. Rev.} {\bf D24} (1981) 2959.}
\refdef{thornweeparton}{C.B. Thorn,
{\sl Phys. Rev.} {\bf D19} (1979) 639.}
\refdef{thornfermions}{C. B. Thorn,
{\sl Phys. Rev.}, {\bf D4} (1971) 1112.}
\refdef{thorncop}{C. B. Thorn,
``Calculations in Perturbative String Field Theory'',
in {\it Perspectives in String Theory},  ed. P. diVecchia
and J. L. Petersen,
World Scientific (1988).}
\refdef{thornfishnet}{C.B. Thorn,
{\sl Phys. Rev.} {\bf D17} (1978) 1073.}
\refdef{thornfock}{C.B. Thorn,
{\sl Phys. Rev.} {\bf D20} (1979) 1435.}
\refdef{thorngf}{C. B. Thorn,
{\sl Nucl. Phys.} {\bf B287} (1987) 61.}
\refdef{thornhotqcd}{C. B. Thorn,
{\sl Phys. Lett.} {\bf 99B} (1981) 458.}
\refdef{thornkacdet}{C. B. Thorn,
{\sl Nucl. Phys.} {\bf B248} (1984) 551.}
\refdef{thornlattice}{C. B. Thorn,
``Lattice Strings'',
in {\it Strings '88}, ed. S. J. Gates, Jr.,
C. R. Preitschopf, and W. Siegel, World Scientific
Publishing Co. (1989).}
\refdef{thornlcft}{C. B. Thorn,
{\sl Nucl. Phys.} {\bf B263} (1986) 493.}
\refdef{thornmosc}{C. B. Thorn,
``Reformulating String Theory with the 1/N Expansion,''
Invited talk at the First International A. D. Sakharov
Conference on Physics, 27--31 May, 1991,
Lebedev Institute, Moscow.}
\refdef{thornnploop}{C. B. Thorn,
{\sl Phys. Rev.} {\bf D2} (1970) 1071.}
\refdef{thornphysstate}{C. B. Thorn,
{\sl Nucl. Phys.} {\bf B286} (1987) 61.}
\refdef{thornreview}{C. B. Thorn,
{\sl Physics Reports} {\bf 175} (1989) 1-101,
and references cited therein.
}
\refdef{thornsantab}{C. B. Thorn, in
{\it Unified String Theories,} ed. M. Green and D. Gross,
World Scientific Publishing Co. (1986).}
\refdef{thorntrieste}{C. B. Thorn,
``Lectures on String Theory'', in {\it Superstrings '88},
Proc. of the Trieste Spring School, M. Green, M. Grisaru
R. Iengo, and A. Strominger, World Scientific Publishing Co. (1988).}
\refdef{thornsubdual}{C. B. Thorn,
{\sl Phys. Lett.} {\bf 242B} (1990) 364-370.}
\refdef{thornvertex}{C. B. Thorn,
in {\it Vertex Operators in Mathematics and Physics,
Proc. 1983 MSRI Conference}, ed. J. Lepowsky {\it et al}
(Springer-Verlag, New York, 1985) 411.}
\refdef{tyutin}{I. V. Tyutin,
Lebedev preprint FIAN No. 39 (1975).}
\refdef{vafaw}{C. Vafa and E. Witten,
{\sl Phys. Rev. Lett.} {\bf 53} (1984) 535.}
\refdef{vandamv}{H. Van Dam and M. Veltman,
{\sl Nucl. Phys.} {\bf B22} (1970) 397.}
\refdef{veneziano}{G. Veneziano,
{\sl Nuovo Cim.} {\bf 57A} (1968) 190.}
\refdef{verlindev}{E. Verlinde and H. Verlinde,
{\sl Phys. Lett.} {\bf 192B} (1987) 99;
{\sl Nucl. Phys.} {\bf B288} (1987) 357.}
\refdef{verlindefuse}{E. Verlinde,
{\sl Nucl. Phys.} {\bf B300} (1988) 360.}
\refdef{virasoroamp}{M. Virasoro,
{\sl Phys. Rev.} {\bf 177} (1969) 2309.}
\refdef{virasorogen}{M. Virasoro,
{\sl Phys. Rev.} {\bf D1} (1970) 2933.}
\refdef{wendt}{Wendt, C.,
{\sl Nucl. Phys.} {\bf B314} (1989) 209.}
\refdef{weinbergw}{S. Weinberg and E. Witten,
{\sl Phys. Lett.} {\bf 96B} (1980) 59.}
\refdef{weis}{J. H. Weis, unpublished
private communication as cited in
R. C. Brower and C. B. Thorn,
{\sl Nucl. Phys.} {\bf 31} (1971) 163.}
\refdef{weyl}{H. Weyl,
{\it The Classical Groups}, Princeton Univ. Press,
Princeton, NJ, 1946.}
\refdef{whittaker}{Whittaker, E. T. and G. N. Watson,
``A Course in ModernAnalysis'',
Cambridge Univ. Press (1927) \S 20$\cdot $14.}
\refdef{wittensup}{E. Witten,
{\sl Nucl. Phys.} {\bf B276} (1986) 291.}
\refdef{witten}{E. Witten,
{\sl Nucl. Phys.} {\bf B268} (1986) 253.}
\refdef{witten3dgrav}{E. Witten,
{\sl Nucl. Phys.} {\bf B311} (1988-9) 46.}
\refdef{wittenjones}{E. Witten,
{\sl Comm. Math. Phys.} {\bf 121} (1989) 351.}
\refdef{wittentop}{E. Witten,
{\sl Comm. Math. Phys.} {\bf 117} (1988) 353.}
\refdef{woodards}{M. Srednicki and R. Woodard,
{\sl Nucl. Phys.} {\bf B293} (1987) 612.}
\refdef{woodard}{R. Woodard, Brown University preprint,
BROWN HET-652 (1988).}
\refdef{woodardunstablesft}{D. A. Eliezer and R. P. Woodard,
{\sl Nucl. Phys.} {\bf B325} (1989) 389.}
\refdef{yamron}{J. P. Yamron,
{\sl Phys. Lett.} {\bf 174B} (1986) 69.}
\refdef{yamrondeltarep}{J. P. Yamron,
{\sl Phys. Lett.} {\bf 187B}(1987)67.}
\refdef{zeeinducedgrav}{A. Zee,
{\sl Phys. Rev. Lett.} {\bf 42} (1979) 417;
{\sl Phys. Rev.} {\bf D23} (1981) 858;
{\sl Phys. Rev. Lett.} {\bf 48} (1982) 295;
{\sl Phys. Lett.} {\bf 109B} (1982) 183.}
\refdef{zeldovich}{Ya. B. Zel'dovich,
{\sl Zh. Eksp. Teor. Fiz. Pis'ma Red} {\bf 6} (1967) 883
[{\sl JETP Lett} {\bf 6} (1967) 316].}
\refdef{zwiebach}{B. Zwiebach,
``Constraints on Covariant Theories for
Closed String Fields,''
MIT preprint (1988), CTP \#1583;
``A Note on Covariant Feynman Rules for
Closed Strings,''
MIT preprint (1988), MIT-CTP-1598 (1988);
``Closed String couplings with Modular Parameters,''
MIT preprint (1988), CTP \#1634.}
\refdef{zembaz}{C. Zemba and B. Zwiebach,
``Tadpole Graph in Covariant Closed String Field
Theory,''
MIT preprint (1988), MIT-CTP-1633.}
\refdef{zembaz}{C. Zemba and B. Zwiebach,
``Tadpole Graph in Covariant Closed String Field
Theory,''
MIT preprint (1988), MIT-CTP-1633.}
\refdef{zwiebachclosed}{B. Zwiebach,
``Constraints on Covariant Theories for
Closed String Fields,''
MIT preprint (1988), CTP \#1583;
``A Note on Covariant Feynman Rules for
Closed Strings,''
MIT preprint (1988), MIT-CTP-1598 (1988);
``Closed String couplings with Modular Parameters,''
MIT preprint (1988), CTP \#1634.}
\refdef{nonpolyclosed}{
M. Saadi and B. Zwiebach, {\sl Ann. Phys.} {\bf 192} (1989) 213;
H. Sonoda and B. Zwiebach, MIT preprints MIT-CTP-1774, 1758;
B. Zwiebach, MIT preprint MIT-CTP-1787;
T. Kugo, H. Kunitomo and K. Suehiro, {\sl Phys. Lett.} {\bf B226} (1989) 48;
T. Kugo and K. Suehiro, Kyoto University preprint KUNS-988.}
\refdef{freegs}{S.J. Gates, M.T. Grisaru, U. Lindstr\"om, M. Ro\v cek,
W. Siegel, P. van Nieuwenhuizen and A.E. van de Ven, {Phys. Lett.}
{\bf B225} (1989) 44; U. Lindstr\"om, M. Ro\v cek,
W. Siegel, P. van Nieuwenhuizen and A.E. van de Ven, Stony Brook preprint
ITP-SB-89-38;
R.E. Kallosh, {Phys. Lett.}
{\bf B224} (1989) 273,
{\bf B225} (1989) 49;
M.B. Green and C.M. Hull,
{Phys. Lett.}
{\bf B225} (1989) 57.}
\refdef{wronggs}{J.M.L. Fisch and M. Hennaux, Universit\'e Libre de Bruxelles
preprint ULB TH2/89-04;
F. Bastianelli, G.W. Delius and E. Laenen, Stony Brook preprint ITP-SB-89-51;
U. Lindstr\"om, M. Ro\v cek,
W. Siegel, P. van Nieuwenhuizen and A.E. van de Ven, Stony Brook preprint
ITP-SB-89-76.}
\refdef{openliouville}{O. Alvarez, {\sl Nucl. Phys.} {\bf B216}
(1983) 125;
B. M. Barbashov and V. V. Nesterenko, {\sl Theor. Math. Phys.}
{\bf 56} (1984) 752;
B. Durhuus, H. B. Nielsen, P. Olesen, and J. L. Petersen,
           {\sl Nucl. Phys.} {\bf B196} (1982) 498;
B. Durhuus, P. Olesen, and J. L. Petersen,
           {\sl Nucl. Phys.} {\bf B198} (1982) 157, {\bf B201} (1982) 176;
L. Johansson, A. Kihlberg, and R. Marnelius, {\sl Phys. Rev.}
{\bf D29} (1984) 2798;   
P. Mansfield, {\sl Phys. Rev.} {\bf D28} (1983) 391, {\sl Nucl. Phys.}
   {\bf B222} (1983) 419;    
E. Onofri and
M. A. Virasoro, {\sl Nucl. Phys.} {\bf B201} (1982) 159.}
\def\frac#1#2{{#1 \over #2}}
\def\goesas#1#2{\ \ {\phantom{a} \atop
                \widetilde{#1\rightarrow #2}}}

\def\doeack{\foot{Work supported in part by the Department of Energy,
      contract DE--FG05--86ER--40272.}}
%
\Pubnum = {UFIFT-92-12\cr
}
\date = {}
\titlepage
\title {Quark-Antiquark Regge Trajectories in Large $N_c$ $QCD$\doeack}
\author{Michael McGuigan and Charles B. Thorn}
\address{Department of Physics\break
University of Florida, Gainesville, FL 32611}
\abstract
We apply methods developed by Lovelace, Lipatov, and Kirschner to
evaluate the leading Regge trajectories $\alpha(t)$ with the quantum numbers
of nonexotic quark-antiquark mesons at $N_c=\infty$
in the limit $t\rightarrow-\infty$
where renormalization group improved perturbation theory should be
valid. We discuss the compatibility of nonlinear trajectories with
narrow resonance approximations.\endpage
\pagenumbers

It is unlikely that Quantum Chromodynamics, the consensus theory of
strong interactions, can be exactly solved with realistic values
for all parameters. However, asymptotic freedom allows the
application of weak coupling techniques such as perturbation
theory to obtain
the predictions of $QCD$ for processes controlled by short
distance dynamics. Besides high momentum transfer collision
phenomena, one can hope to use such weak coupling techniques
for computing the mass spectrum of
hadrons containing only very heavy quarks. But for hadrons
containing light quarks and also for glueballs, strong coupling dynamics
is unavoidable.

It is reasonable to first confront these strong coupling
issues making as many simplifying idealizations as possible.
Thus we might idealize light quarks to
massless quarks and delete the heavy quarks all together.
Since $m_u,m_d<<\Lambda_{QCD}$ this idealized theory
should give an excellent approximation to the dynamics of
up and down quarks, which is to say all ordinary
hadronic matter. These
idealizations leave us with a theory with no further
free parameters.
Unfortunately, making the
quarks massless does not simplify the dynamics enough for an
analytical treatment: the $S$-matrix is non-trivial in all channels
including those with particle production. Furthermore,
the bound state spectrum includes all nuclei as well
as the lowest mass hadron in each flavor sector.

That is why `t Hooft's idea of exploiting the $N_c\rightarrow\infty$
limit\[thooftlargen] is so attractive. In this limit the scattering
amplitudes involving
hadrons vanish and in lowest nonvanishing order are meromorphic
in the channel invariants, just as the tree approximation to
a quantum field theory. Nor do the nuclei bind in this limit.
Thus an exact solution in this limit really would be significantly simpler
than the exact solution at $N_c=3$.
Even if
the definitive tests of $QCD$ must come from large scale computation,
a successful analytical understanding of all hadronic phenomena to $30\%$
accuracy would be very desirable. Unfortunately, to date efforts
to evaluate the large $N_c$ limit have failed: at least with
available methods the limiting theory seems almost as intractable as
the finite $N_c$ theory. Nonetheless,
we think it is worthwhile to develop as much insight into the
nature of the hoped for solution as possible.

String theory started as
an effort to build exactly the sort of approximation to strong
interaction dynamics that is provided by large $N_c$ $QCD$.
Since that approach led to the ``wrong'' answer, we should
understand how the expected properties of $QCD$
are different from those of string theory. In string theory
the Regge trajectory functions $\alpha_{string}(t)=\alpha^\prime t
+\alpha_0$ where $\alpha^\prime=1/2\pi T_0$ with $T_0$ the rest
tension in the string, play a central role in the string scattering
amplitudes: they appear directly as the arguments of the
gamma functions which characterize the Veneziano four string
function $A_4(s,t)=g^2_{string}
\Gamma(-\alpha(s))\Gamma(-\alpha(t))/\Gamma(-\alpha(s)-\alpha(t))$.
The meromorphy of $A_4$ in $s$ and $t$ follows directly from
that of the gamma functions and the exact linearity of the Regge
trajectories. Since the large $N_c$ hadron amplitudes are also
meromorphic in $s$ and $t$, the trajectory functions themselves
should be good characteristics of the similarities and differences
between string theory and $QCD$. Also they might carry some hints
about the solution of large $N_c$ $QCD$.

In this letter we
study the Regge trajectories of large $N_c$ $QCD$ in the
``meson'' channels (\ie\ those interpolating the rotational
states of quark-antiquark mesons), in the limit of large
negative $t$ where perturbative $QCD$ should be applicable. We
follow ideas and methods developed by Lovelace\[lovelacefreeregge],
Lipatov\[lipatovpomeron], and
Kirschner and Lipatov\[kirschnerlbarereggeons]. These methods are essentially
renormalization group improved calculations based on summing
leading logarithmic contributions of Feynman graphs. Such
methods can only give the trajectory functions in the
weak coupling approximation. Since the coupling
``runs'' with the scale $\lambda(-t)\equiv N_cg^2_s(-t)/4\pi^2
\sim 12/11\ln(-t/\Lambda^2_{QCD})$ this means that we can obtain only the
large negative $t$ behavior of the Regge trajectories using
these methods. \foot{Note that the large positive $t$ behavior of the
trajectories is characterized
by the confining force and should be asymptotically linear
$\alpha_{QCD}(t)\sim t/2\pi k$ where $kR$ is the confining
term in the $q\bar q$ interaction energy.}

Our first task is to identify the leading logarithmic contributions
to a scattering process involving the exchange of a $q\bar q$
pair. In any gauge theory the ladder diagrams, which iterate
gauge boson exchange between two fermion lines, contribute
two powers of $\ln(s/\mu^2)$ for each additional rung. Thus
the leading logarithms are actually double logarithmic and
dominate the single logarithms of renormalization: the leading
log sums will therefore {\it not} include running coupling
effects. Thus we proceed in two
steps. First we evaluate the amplitudes to double logarithmic
accuracy and then incorporate renormalization effects which
make the coupling run in the second step. As shown
in 1967\[gorshkovglf] for $QED$,
the first step typically leads to a fixed square root branch point
in the angular momentum plane. For $QED$ processes
involving the exchange of total zero charge, the leading
double logs come only from the ladder sum, which
produces a branch point
located at $J=\sqrt{2\alpha/\pi}$ where $\alpha\approx 1/137$ is the fine
structure constant. When non zero charge is exchanged, there
are additional double logarithmic contributiions coming from
soft ``bremstrahlung'' photons which either form crossed
rungs in the ladder or Sudakov vertex corrections.
Similarly, in $QCD$, as discussed by
Kirschner and Lipatov\[kirschnerlnonvac],
the double logarithmic contributions of the basic ladder
diagrams of Fig. 1 are supplemented by soft gluon bremstrahlung
graphs. Fortunately for us, these additional
diagrams are nonleading in the $1/N_c$ expansion when the
ladder structure is ``hooked on'' to color singlet hadron
vertices\foot{Note that these hadron ``form factors,''
represented by the left and right hand parts of Fig. 1,
involve on-shell mesons and cannot be calculated perturbatively.}
and so do not enter into our
calculation of the Regge trajectories of
large $N_c~QCD$. Thus the double logarithmic sum is identical
to the $QED$ zero charge exchange case with $\lambda\equiv
N_c g_s^2/4\pi^2$ substituted for $2\alpha/\pi$.

Since renormalization effects are neglected in the leading
double logarithmic approximation, the results depend on a
fixed coupling constant $\lambda$. Asymptotic freedom
must at least make the location of the
singularities in the angular momentum plane vary with
$t$ according to the replacement $\lambda\rightarrow\lambda(-t)$,
but actually
the cut is changed into a distribution
of Regge poles accumulating at 0 as $t\rightarrow-\infty$.
This phenomenon was first uncovered by
Lovelace\[lovelacefreeregge] for the
case of $\phi^3$ theory in 6 space-time dimensions where the
accumulation point is at $J=-1$. He
analysed the Bethe-Salpeter (B-S) equation with a kernel improved
to include the effects of asymptotic freedom. This
equation produced partial wave amplitudes with
only pole singularities in the angular momentum plane. Since he
only considered the case $t=0$, his
results for the pole locations (Regge intercepts) were
untrustworthy. (The low momentum theory is a strong
coupling one and the B-S equation
is not valid there.) This shortcoming was removed by
Kirschner and Lipatov\[kirschnerlbarereggeons],
who
incorporated $t$ dependence
in leading order and obtained $\alpha(t)$ for
large negative $t$ instead of $\alpha(0)$. With large enough
$t$, the effective coupling is weak, justifying
the B-S equation.
Earlier, Lipatov\[lipatovpomeron]
had obtained similar results for the asymptotic
behavior of the pomeron (glueball) trajectory in $QCD$.
In this note we find
the corresponding asymptotic behavior of
the $q\bar q$ trajectories in large $N_c~QCD$.

In order to incorporate the running coupling,
we consider the B-S equation which
sums the ladder subgraphs in Fig. 1. We represent the
Green's function for the ladder
subgraphs as a matrix $\Psi_{ab}$ in the Dirac indices
of the $q\bar q$ lines coming in at the left. We apply the Dirac
operators to these two lines in the coordinate
representation to obtain (for simplicity we take all $m_q=0$)
$$\gamma\cdot\partial_1\Psi(x_1,x_2;y_1,y_2)\gamma\cdot\overleftarrow
\partial_2= -\delta(x_1-y_1)\delta(x_2-y_2)
+\lambda(x_{12}^{-2})\gamma^\mu\Psi\gamma^\nu d_{\mu\nu}
(x_{12})$$
where we have followed Lovelace's treatment
of $\phi^3_6$, replacing the coupling
constant by the running coupling $\lambda(x^{-2})\approx
-12/11\ln(x^2\Lambda_{QCD}^2)$.
In a general covariant gauge we define the coordinate space
propagator by $d_{\mu\nu}(x)/{4\pi^2}$ with
$$
d_{\mu\nu}(x)
=-i\int{d^4p\over(2\pi)^2}e^{ix\cdot p}{\eta_{\mu\nu}-\zeta p_\mu p_\nu/p^2
\over p^2-i\epsilon}=
{(1+\zeta)\eta_{\mu\nu}\over 2x^2}+{(1-\zeta)x_\mu x_\nu\over x^4}.
$$
The B-S equation is not gauge invariant, but
violations of gauge invariance will be small for weak
coupling. Thus if we only keep leading order answers,
our results should be gauge invariant. We keep $\zeta$
arbitrary so we can confirm this.
We expect this equation to be accurate when $\lambda<<1$
\ie\ for $x_{12}^2\Lambda^2_{QCD}<<1$. The singularities in
the $t$ channel are controlled by the solutions of the
homogeneous equation
$$\gamma\cdot\partial_1\Psi(x_1,x_2)\gamma\cdot\overleftarrow
\partial_2= \lambda( x^{-2}_{12})
\gamma^\mu\Psi(x_1,x_2)\gamma^\nu d_{\mu\nu}
(x_{12}).$$
It is convenient to work with $CM$ and relative coordinates
$r=(x_1+x_2)/2$ and $\rho=x_1-x_2$ and to Fourier transform
with respect to $r$, whose conjugate variable is q so
that $q^2=-t$. Then the homogeneous equation
reads
$$
\gamma\cdot({\partial\over\partial\rho}-{iq\over2})
\tilde\Psi(\rho,q)\gamma\cdot(-{\overleftarrow
\partial\over\partial\rho}-{iq\over2})=
\lambda(\rho^{-2})\gamma^\mu\tilde\Psi(\rho,q)\gamma^\nu d_{\mu\nu}
(\rho).$$
We can only make use of this equation for small $\rho$
and large $q$ when the effective coupling associated with
both scales in the problem is small.

For $q\rho<<1$, but $q, \rho^{-1}>>\Lambda_{QCD}$,
the B-S equation becomes quite
manageable. This limit reduces it to the $q=0$ case, with
its $O(4)$ symmetry. As in \Refer{kirschnerlbarereggeons}
one can conveniently consider the $\ell^{th}$ partial waves
in this limit by making the ansatz
$$\tilde\Psi_\ell(\rho)={(\rho\cdot\xi)^{\ell-1}\over|\rho|^\ell}
\left[{\rho\cdot\gamma\rho\cdot\xi\over\rho^2}f(|\rho|)
+\xi\cdot\gamma g(|\rho|)\right],\(partialwave)$$
with $\xi^\mu$ a fixed light-like four-vector, so that one is
forming traceless symmetric tensors of rank $\ell$. Plugging
this ansatz into the B-S equation then yields the pair of equations
$$\eqalign{
[({\cal D}-\ell)^2-4]f+2[{\cal D}-\ell][{\cal D}-\ell-2]g
=&\lambda(\rho^{-2})[\zeta f-(1-\zeta)g]\cr
2\ell[{\cal D}+1]f-[{\cal D}-\ell]^2g=&\lambda(\rho^{-2})g.\cr}
\(smallrhoeq)
$$
Here ${\cal D}\equiv|\rho|\partial/\partial|\rho|$ $=\partial/\partial
\ln(|\rho|\Lambda)$ is simply the
scaling derivative in the magnitude of $\rho$.

Eq.\(smallrhoeq) is a spinor version of the small $\rho$ B-S
equation for the $\phi^3_6$ theory analyzed in
\Refer{lovelacefreeregge,kirschnerlbarereggeons}. With
$\lambda(\rho^{-2})$ approximated by $-6/11\ln(|\rho|\Lambda)$, the
Laplace transforms of $\lambda f$, $\lambda g$ satisfy a
pair of first order differential equations
in ${\cal D}\equiv -2i\hat\nu$. Unlike the single equation in the
$\phi^3_6$ case which can immediately be integrated, this pair of
equations is equivalent to a single component second order
equation which cannot be so readily solved. However, the
leading angular momentum singularity (with largest Re $\ell$) is
controlled (for small $\lambda$) by $\ell$ and $\hat\nu$ small compared
to unity. This is made clear by considering the determinant of
the coefficient matrix on the l.h.s. of these equations:
$$\det\pmatrix{(-2i\nu-\ell)^2-4&2[-2i\nu-\ell][-2i\nu-\ell-2]\cr
2\ell[-2i\nu+1]&-[-2i\nu-\ell]^2\cr}=-[\ell^2+4\nu^2][(\ell+2)^2+4\nu^2].$$
In the small $\rho,\nu$ limit, the second of \(smallrhoeq)
shows
that $2\ell f\approx(\lambda(\rho^{-2})+[{\cal D}-\ell]^2)g$. Inserting
this approximate form for $f$ into the first equation and making the
same approximations there gives
$(-{\cal D}^2+\ell^2)g\approx \lambda(\rho^{-2})g$
the Laplace transform of which gives a first order equation
in $\nu$.
Thus we see that this leading Regge singularity is controlled by
equations independent of the gauge $\zeta$, as we anticipated for
weak coupling. Looking back to the full gauge dependent small
$\rho$ equations, we notice that in Landau gauge ($\zeta=0$)
one can eliminate $f$ in favor of $g$ in a $\rho$ independent
way. Thus in this gauge the Laplace transformed equations are
first order in $\nu$ and can be directly integrated.

We exploit this simplification by setting
$\zeta=0$ in the following.
Solving the second equation for $f$ and substituting in the
first we obtain
$$\eqalign{
[\ell^2-{\cal D}^2][(\ell+2)^2-{\cal D}^2]g=&[4-{\cal D}^2-\ell^2-2\ell]
\lambda(\rho^{-2})g.\cr}
\(smallrholandau)
$$
The equation for $g$ is now quite similar to the $\phi^3_6$ case and
we can repeat the steps in \Refer{kirschnerlbarereggeons} to
derive the asymptotic behavior of the Regge trajectories. The
equation for $g$ is first solved by writing it in the form
$$-\ln(|\rho|^2\Lambda^2)(\lambda(\rho^{-2})g)=
{12\over11}{4+4\hat\nu^2-\ell^2-2\ell
\over[\ell^2+4\hat\nu^2][(\ell+2)^2+4\hat\nu^2]}(\lambda(\rho^{-2})g),$$
where we have put ${\cal D}=-2i\hat\nu$. Since
$[\hat\nu,-\ln(\rho^2\Lambda^2)]=-i$ we can interpret
$R\equiv -\ln(\rho^2\Lambda^2)\rightarrow i\partial/\partial\hat\nu$,
integrate the equation and then transform back to the coordinate
representation to obtain
$$\lambda(\rho^{-2})g=\int_{-\infty}^\infty d\nu\exp\left\{
i\nu R-{12i\over11}\int_0^\nu d\nu^\prime{4+4\nu^\prime-\ell^2-2\ell
\over[\ell^2+4\nu^{\prime2}][(\ell+2)^2+4\nu^{\prime2}]}\right\}.
\(runningsolution)$$
This solution of the small $\rho$ B-S equation is, in fact,
a solution of the full B-S equation for $q=0$.
But of course the B-S equation
is only a good approximation for large $q$ and small $\rho$.
In \Refer{kirschnerlbarereggeons} the analogous
solution for the $\phi^3_6$ theory is used to gain information
about the large $q$ behavior of the trajectories by noting that for
$\rho q<<1$ but $\rho$ not too small one can have $\lambda(\rho^{-2})
\approx\lambda(q^2)$ for a large range of $\rho$ (essentially because the
scale dependence of $\lambda$ is only logarithmic). Thus instead of
requiring regularity of the solution at $R=0$ as
in \Refer{lovelacefreeregge},
the solution is matched to that
of the B-S equation with a $\rho$ independent coupling
taken to be $\lambda(q^2)<<1$. For consistency of the weak coupling
approximation this matching must be imposed at large $R$ (small $\rho$).

The large $R$ behavior of \(runningsolution) is exponentially
damped for extremely large $R$ but there is oscillatory behavior
for $R$ not too large. This can be extracted by finding the saddle
points
$$\nu_0^2={3\over22R}-{1+\ell\over2}-{\ell^2\over4}
\pm\sqrt{\left[{3\over22R}-{1+\ell\over2}-{\ell^2\over4}
\right]^2+{3(4-\ell^2-2\ell)\over44 R}
-{\ell^2(\ell+2)^2\over16}}.\(saddlesquare)$$
We see that there are two real values of $\nu_0$ provided
$R<{12\over11}\left[{4-\ell(\ell+2)\over\ell^2(\ell+2)^2}\right]$
which is consistent with large $R$ provided $|\ell|<<1$.
In this regime, $\lambda g$ is well approximated by the
saddle point evaluation
$$\lambda g\approx 2\sqrt{6\pi\over11\Phi^{\prime\prime}(\nu_0)}
\cos\left[{\pi\over4}+\nu_0R-{12\over11}\Phi(\nu_0)\right]\(saddleg)$$
where $\nu_0$ is the positive square root of \(saddlesquare)
taken with the plus sign, and
$$\Phi(\nu_0)=
{2-\ell^2-\ell\over8\ell(\ell+1)}i\ln{1-2i\nu_0/\ell\over
1+2i\nu_0/\ell}
+{\ell(\ell+3)\over8(\ell+1)(\ell+2)}i\ln{1-2i\nu_0/(\ell+2)\over
1+2i\nu_0/(\ell+2)}.
$$
For sufficiently large $R$, the $R$ dependence of $\nu_0$ can be
neglected and \(saddleg) can be matched to the small $\rho$
solution of the B-S equation with constant coupling $\lambda(q^2)$.
Scale invariance of the finite $q$ B-S equation implies that
the solution is a function of $\rho|q|$. In the
small $\rho$ limit it therefore has the behavior
$$C[(\rho|q|)^{-2i\nu_1}+(\rho|q|)^{+2i\nu_1}e^{i\delta(\ell,\nu_1)}]
=2Ce^{i\delta/2}\cos\left[{\delta\over2}+\nu_1\ln|\rho|^2|q|^2\right]
\(constantcoupling)$$
where $\nu_1$ is $\nu_0$ with $12/11R$ replaced by $\lambda(q^2)$.

Replacing $\nu_0$ by $\nu_1$ in \(saddleg) and comparing to
\(constantcoupling) gives the matching condition
$$
\lambda(q^2)\Phi(\nu_1)-\nu_1={1\over\ln(q^2/\Lambda^2)}
\left\{r\pi+{\delta\over2}+{\pi\over4}\right\}
\(matching)
$$
where $r=$ integer.
We shall find that the consistent small coupling solution of
these equations gives $\ell=O(\lambda^{1/2})$ and $\nu_1=O(\lambda^{2/3})$.
Thus $\nu_1/\ell^2=O(\lambda^{-1/3})$ and $\nu^3/\ell^4=O(1)$.
Neglecting all terms that vanish at zero $\lambda$ in $\Phi(\nu_1)$,
the matching condition simplifies to
$$\lambda(q^2)\left({\nu_1\over\ell^2}-{4\nu^3_1\over3\ell^4}\right)-\nu_1
  ={1\over\ln(q^2/\Lambda^2)}
\left\{r\pi+{\delta\over2}+{\pi\over4}\right\}$$
With these approximations $\lambda\approx\ell^2(1+{4\nu_1^2/\ell^2})$
so the leading term on the l.h.s. cancels and we are left with
$${8\nu^3_1\over3\ell^2}\approx{1\over\ln(q^2/\Lambda^2)}
\left\{r\pi+{\delta\over2}+{\pi\over4}\right\}$$
Replacing $\ell^2$ by $\lambda(q^2)$, we thus find the Regge trajectory
asymptotics
$$\eqalign{
\alpha_r(t)&\goesas t{-\infty}
\sqrt{\lambda(-t)-4\nu_1^2}\cr
&\approx
\sqrt{\lambda(-t)}
\left(1-2\left[{11\over32}\right]^{2/3}\lambda^{1/3}(-t)\left(
r\pi+{\delta\over2}+{\pi\over4}\right)^{2/3}+\cdots\right).\cr}
\(rhotrajectory)$$
Notice that we have an infinite number of trajectories
accumulating at 0 in the limit $t\rightarrow-\infty$, with almost
identical behavior to those of the $\phi^3_6$ theory obtained in
\Refer{kirschnerlbarereggeons}.

The phase $\delta$ is not determined from the small $\rho$
dynamics considered so far. It must be determined
by the dynamics at $\rho q=O(1)$. However, except in exceptional
cases, $\delta=\pi$ in the limit we are considering.
This is because this limit involves $\nu\approx0$ in
\(constantcoupling). For $\nu=0$ the two behaviors $\rho^{\pm2i\nu}$
 are replaced by 1 and $\ln\rho$. Generically, both behaviors
will be present, and unless the coefficient of $\ln\rho$ exactly
vanishes, the behavior for $\nu$ slightly different from zero
must be
$$N\left({c_1\over\nu}[(q\rho)^{-2i\nu}-(q\rho)^{2i\nu}]+c_2(q\rho)^{-2i\nu}
+c_3(q\rho)^{+2i\nu}\right)$$
with $c_a$ finite at $\nu=0$. If the $\ln\rho$ term is present at
$\nu=0$, then $c_1\neq0$ there and $\delta=\pi$ at $\nu=0$. This
is analogous to the generic vanishing of phase shifts at zero
energy. In that analogy the case $c_1=0$ corresponds to a
``zero energy resonance.'' For the $\phi^3_6$
case, $\delta=\pi$ was shown by explicit
solution of the constant coupling B-S equation
using conformal invariance\[kirschnerlbarereggeons].
In gauge theories, a conformal transformation changes the
gauge condition, so the B-S equation, being gauge noninvariant,
is scale invariant but not conformally covariant.
Lacking an explicit solution, we can only state that it
is likely, but not proven, that $\delta=\pi$ for large $N_c~QCD$.

We close with some comments about the significance of nonlinear
Regge trajectories for large $N_c~QCD$. There is a common
belief\[petrovs] that narrow resonance approximations require
exactly linear (or at worst polynomial) Regge trajectories.
However this conclusion depends on a maximal analyticity
assumption that the
trajectory functions are free of singularities in the $t$ plane
cut on the right at threshold branch points\[mandelstambrandeis].
Since ${\rm Im}\ \alpha(t)$, the discontinuity
across the threshold cut, is proportional to the resonance
widths, the trajectories would then be entire functions in
the limit of zero width resonances.
We have seen that the $q\bar q$ trajectories of large $N_c~QCD$
approach constants as $t\rightarrow-\infty$, and confinement
together with infinite $N_c$
implies linear behavior as $t\rightarrow+\infty$ as well as no threshold
branch points. The inescapable conclusion is that the maximal
analyticity assumption fails for the Regge trajectories of infinite $N_c~QCD$
and there are additional singularities in the $t$ plane. This
is probably also true at $N_c=3$
 since there is no
good physical basis for the absence of additional singularities.
In \Refer{thornnonlinear} (for earlier models see also \Refer{coon,coonb})
one of us discussed some examples of
narrow resonance models with nonlinear trajectories with algebraic
branch points in the complex $t$ plane. Some of these models
have trajectories which are asymptotically linear for large
positive $t$ and approach constants at large negative $t$. Unfortunately,
they approach these constants as an inverse power of $t$ rather
than an inverse power of $\ln(-t)$, so they are not candidates for
large $N_c~QCD$. Nonetheless, they do show that nonlinear trajectories
are compatible with narrow resonances, and indicate a direction
toward solving large $N_c~QCD$.
\subsection{Acknowledgements} We should like to thank Al Mueller for
valuable discussions on the Regge limit of $QCD$.

\vskip18pt
\titlestyle{References}\nobreak
\reflist{}
\vskip18pt
\subsection{Figure Caption} A typical large $N_c$ diagram contributing
to meson scattering with the exchange of a $q\bar q$ ladder structure.
The leading log approximation as $s\rightarrow\infty$ is the sum of
graphs with an arbitrary number of gluon rungs represented by the
vertical double lines.
\bye